\documentclass{emulateapj}
  \usepackage{graphicx}
  \usepackage{epstopdf}
\usepackage{amssymb}
\usepackage{apjfonts}
\usepackage{natbib}
\usepackage{subfigure}

\usepackage{hyperref}


\long\def\comment#1{}

\def\W2{{\cal W}}

\newcommand{\ident}{\ensuremath{\mathbb{I}}}

\def\be{\begin{equation}}
\def\ee{\end{equation}}
\def\bea{\begin{eqnarray}}
\def\eea{\end{eqnarray}}

\def\cmm2{{\,\rm cm^{-2}}}
\def\cm2{{\,{\rm cm}^2}}
\def\cmm3{{\,{\rm cm}^{-3}}}
\def\gcmm3{{\,{\rm g\,cm^{-3}}}}

\def\fun#1#2{\lower3.6pt\vbox{\baselineskip0pt\lineskip.9pt
  \ialign{$\mathsurround=0pt#1\hfil##\hfil$\crcr#2\crcr\sim\crcr}}}

\shortauthors{Schneider, Holm \& Knox}
\shorttitle{Intelligent Design}

\begin{document}
%\DeclareGraphicsExtensions{.pdf,.png,.jpg,.gif}

\title{Intelligent Design: On the Emulation of  Cosmological Simulations} 

\author{Michael D. Schneider}
\affil{Institute for Computational Cosmology,
Department of Physics,
Durham University,\\
South Road, Durham DH1 3LE, UK.}
\email{michael.schneider@durham.ac.uk}
\and
\author{\'{O}skar Holm and Lloyd Knox}
\affil{Department of Physics,
 1 Shields Avenue,
 University of California, 
 Davis, CA 95616, USA.}

\begin{abstract} Simulation design is the choice of locations in
  parameter space at which simulations are to be run and is the first
  step in building an emulator capable of quickly providing
  estimates of simulation results for arbitrary locations in the
  parameter space.  
  We introduce an alteration to the ``OALHS'' design used by
  \citet{heitmann06} that reduces the number of simulation runs
  required to achieve a fixed accuracy in our case study by a factor
  of two.
  We also compare interpolation procedures for emulators and find that
  interpolation via Gaussian Process models and via the much-easier-to-implement 
  polynomial interpolation have comparable accuracy.
  A very simple emulation-building procedure consisting of a design
  sampled from the parameter prior
  distribution, combined with interpolation via polynomials also performs well.
  Although our primary motivation is efficient emulators of non-linear cosmological
  $N$-body simulations, in an Appendix we describe an emulator for the
  cosmic microwave background temperature power spectrum publicly available as computer code.
\end{abstract}


\maketitle

\section{Introduction \label{sec:intro}}  

Many data analysis problems in cosmology require the repeated computation of the
power spectrum, or other summary statistic, expected for a given point
in the model parameter space.  These range from the very expensive, such as
the calculation of weak lensing shear power spectra via ray tracing
through $N$-body simulations, to
the relatively cheap, such as the calculation of cosmic microwave background
(CMB) power spectra from 
Einstein--Boltzmann (EB) equation solvers.  Decreasing the computer resources and
time required for performing these calculations has benefits across this range
of problems.  

The benefits of speed are perhaps most obvious at the more expensive
end of the cost spectrum.  As \citet{heitmann09} have recently argued,
a ``brute-force'' analysis to produce constraints on cosmological
parameters from ``Stage IV'' weak lensing shear power spectra\footnote{Such as
from the Large Synoptic Survey Telescope (LSST) or a future space-based
dark energy mission} would
take a 2048-processor machine 20 years.  Techniques that can
dramatically reduce the computer resource demands are not merely
beneficial, but absolutely necessary.  The ``Cosmic Calibration'' (CC)
framework introduced by \citet{heitmann06} and developed in a series
of papers \citep{habib07,heitmann08,schneider08,heitmann09,lawrence09} has
largely been motivated by this problem. 

Even at the relatively cheap end there are benefits to increased
speed.  Although EB solvers are relatively fast \citep{seljak96}, and can take as
little as tens of seconds on a single processor, much effort has gone
into making power spectrum calculation even faster.   Early work exploited
the angular-diameter distance degeneracy at small angular scales to
reduce pre-compute requirements \citep{tegmark00a,dash02}.  Later work
used more sophisticated training and interpolation procedures, 
and a lot of pre-computing,
 to achieve very high accuracy at very fast post-compute speeds
 \citep{pico,pico2,cosmonet,cosmonet2}.  

Following the terminology of \citet{heitmann06}, by emulator we mean a
machine that can estimate the results of a power spectrum calculation,
without actually doing the calculation.  Typically emulator construction
includes a pre-computing stage in which the calculation of the power spectrum
is performed at some finite set of locations in the parameter space.  The
choice of this set of points (or the ``simulation design'') 
is a critical step in the emulator's construction.  It can have a significant
impact on the number of calculations required to achieve a given accuracy.  
The focus of this paper is on the development of simulation designs that 
minimize the computer resource demands of the pre-compute stage.

The new design technique we introduce includes an application of the
highly efficient ``Orthogonal Array Latin Hypercube Sampling (OALHS)''
which is part of the CC framework.  The essential new elements are the
choice of the parameter basis in which to perform the OALHS and the
volume reduction in the parameter space achieved by constructing
designs in a hypersphere rather than a hypercube.  This
choice is informed by the (appropriately-weighted) use of information
about the rates of change of the power spectrum with respect to
parameter variations.

We also consider the interpolation procedures used for estimating the power spectrum at
arbitrary locations in the parameter space, from those pre-computed
at the design locations.  We compare results from
the Gaussian process (GP) models used in the CC framework with those
from fitting a second-order polynomial.  In all cases we study we find
the (much-simpler-to-implement) second-order polynomial fitting performs
similarly to the GP interpolation.  Their relative merits will vary from
problem to problem however, and we speculate as to when GP will
have advantages.  We find that simulated annealing is an enormously
useful step to include for calibrating the GP parameters.

From our case study we draw conclusions that will be useful for
application to other problems.   Most importantly we
demonstrate that the simulation design, rather than the choice of
an interpolation method, is most critical for emulator construction.   We
also consider the choice of basis functions for reducing the dimensionality
of the statistics to be interpolated, calibration of GP
parameters for interpolation (and how this is affected by the choice of
design algorithm), and the convergence of interpolation
errors as the number of design points is increased.  

As a testing and development ground for our work on emulators we have
chosen to work with CMB temperature power spectra.  Although we are
also motivated by weak lensing applications, the
relative speed of a Boltzmann-code calculation compared to an $N$-body
simulation makes the CMB temperature power spectrum quite convenient
for testing and development.  Accurate calculation of non-linear
matter power spectra is sufficiently expensive as to make their use
for development of emulation strategies a practical
impossibility. \citet{heitmann09} circumvented this problem by
developing and testing their design strategy using an approximate (but
fast) method for calculating the matter power spectra~\citep[\textsc{HALOFIT}][]{smith03}.

While existing CMB emulators are very fast and sufficiently accurate,
they do  have very heavy pre-compute requirements.  The \textsc{PICO} code relies on
tens of thousands of evaluations.  The heavy pre-compute needs make
it difficult to extend an existing emulator's capabilities to include
additional physical effects, such as isocurvature modes or completely
new physical models.  Improved designs can greatly reduce these
difficulties.  As a demonstration we describe and validate 
a publicly available emulator in Appendix~\ref{sec:emuCMB}, 
called \textsc{Emu CMB}, that can 
calculate CMB temperature power spectra out to $\ell = 5000$ 
at sub-percent accuracy as a function of six cosmological 
parameters.  The development of the emulator, utilizing the
efficient design techniques we discuss here, only required
running an EB solver~\citep[CAMB][]{camb}
at 100 design points.  We also show in Appendix~\ref{sec:emuCMB} the
emulator performance for the CMB polarization power spectra, which
still have large errors at low $\ell$.  The large number of training
points required for previous CMB emulators is in part driven by
improving the accuracy for the low-$\ell$ polarization spectra (as
well as covering a larger parameter space).

The structure of this paper is as follows.  We describe each aspect of
our emulator in Section~\ref{sec:emulators_described}.  We lay out the
physical model for CMB power spectra that we use as a case study for the
emulator construction in Section~\ref{sub:cmbspectra}.  We consider 3
simulation design methods that are described in
Section~\ref{sub:simulation_design_construction}.  The basis
decomposition of the simulation design runs and the 2 interpolation
methods we compare are given in Section~\ref{sub:dimreduction} and
\ref{sub:interpolation}.  We present the results of our case study in
Section~\ref{sec:emulator_results} and draw conclusions about the
performance and practicality of the different design and interpolation
methods in Section~\ref{sec:conclusions}.  We present an example of a
CMB temperature power spectrum emulator that meets the requirements for
analysis of modern CMB experiments in Appendix~\ref{sec:emuCMB}.  Code
to construct emulators similar to the one in this section can be
downloaded from \url{http://www.emucmb.info}.  In
Appendix~\ref{sec:coyoteUniverse} we compare the performance of
quadratic polynomial interpolation in the matter power spectrum design
of~\citet{heitmann09} with the GP interpolation errors presented in
their paper.

\section{Emulators Described} \label{sec:emulators_described}
There are three steps to constructing an emulator.
\begin{enumerate}
\item Choose a ``simulation design,'' i.e., the set of points in the parameter space at
which the summary statistic (henceforth assumed to be a power spectrum) will be calculated. 
\item Perform a reduction in the number of dimensions of the power spectrum, 
via an (incomplete) mode decomposition.
\item Specify an interpolation procedure to allow one to estimate the power spectrum that one
would have calculated at {\em any} point in the parameter space.  
\end{enumerate}
Here we review how these three steps are implemented in the ``CC'' work
of \citet{heitmann06,habib07,heitmann08,heitmann09,lawrence09} and also with PICO \citep{pico,pico2}.

\subsection{CMB Temperature Power Spectrum For Our Case Study}
\label{sub:cmbspectra}

The very low expense of CMB temperature power spectrum calculations makes them convenient
to use for testing and development of new emulation techniques.  For this reason,
we pursue emulation of a CMB temperature power spectrum calculator for our case study.  

We set ourselves the goal of building an emulator with sufficient accuracy for analyzing 
the temperature power spectrum constraints that are expected from
Planck\footnote{http://sci.esa.int/planck/}, and to do so in the
simplest possible manner with the fewest possible ``simulations.''  
To inform the emulator construction, we assume prior constraints on
cosmological parameters of similar quality to those from
WMAP5~\citep{nolta09}.  
We consider the following six cosmological parameters:
\begin{equation}
  \boldmath{\theta} = \left(n_s, 100\, \theta^{*}, \ln\left(10^{10} A\right), \omega_c, 
  \omega_b, \tau \right)^{T} ,
\end{equation}
where $n_s$ is the scalar spectral index, $\theta^{*}$ is the angular
size of the sound horizon at last scattering (in radians), $A$ is the
amplitude of the primordial power spectrum defined at the pivot scale
$k=0.002$~Mpc$^{-1}$, $\omega_c\equiv \Omega_c h^2$ is the dark matter
density, $\omega_b\equiv\Omega_b h^2$ is the baryon density, and
$\tau$ is the optical depth.  Our fiducial values for these parameters
are given in the first column of Table~\ref{tb:fidparams} and are
derived from a Markov Chain Monte Carlo (MCMC) chain using the WMAP5
likelihood\footnote{Provided by Alexander van Engelen and Gil Holder}.  
\begin{deluxetable}{cccc}
\tablewidth{0pt}
\tablecaption{\label{tb:fidparams}Fiducial cosmological parameter values
  and 1-$\sigma$ marginal errors derived from the Fisher matrix.}
 \tablehead{
   \colhead{Parameter} & \colhead{Fiducial value} & 
   \colhead{WMAP5 error} & \colhead{Planck error}
 } 
 \startdata
 $n_s$ & 0.959932 & 0.0139280 & 0.0048435\\
  $100\, \theta^{*}$ & 1.039648 & 0.0033038& 0.0003797\\
  $\ln(10^{10} A)$ & 3.049634 & 0.0475892& 0.0346602\\
  $\omega_c$ & 0.1079211 & 0.0069907& 0.0017480\\
  $\omega_b$ & 0.022490 & 0.0006217& 0.0001885\\
  $\tau$ & 0.086429 & 0.0175253& 0.0162973
  \enddata
\end{deluxetable}
We use a Fisher matrix to compute constraints on these six
cosmological parameters given a signal power spectrum computed with
\textsc{CAMB} and a model for the noise power spectrum,
\begin{equation}
  C_{\ell}^{\rm noise} =
  \frac{e^{(\ell\,\sigma_{\rm beam})^2}}{w_{\rm noise}}
\end{equation}
with noise weights $w_{\rm noise} = 1.43\times 10^{14}$ for WMAP5,
$w_{\rm noise} = 4.7\times 10^{16}$ for Planck, beam smearing
$\sigma_{\rm beam}\equiv$~FWHM/2.355, and FWHM~$= 0.00378$ for
WMAP5 (at 90~GHz) and 0.0021 for Planck (at 143~GHz). 

\begin{figure}
  \centering
 \plotone{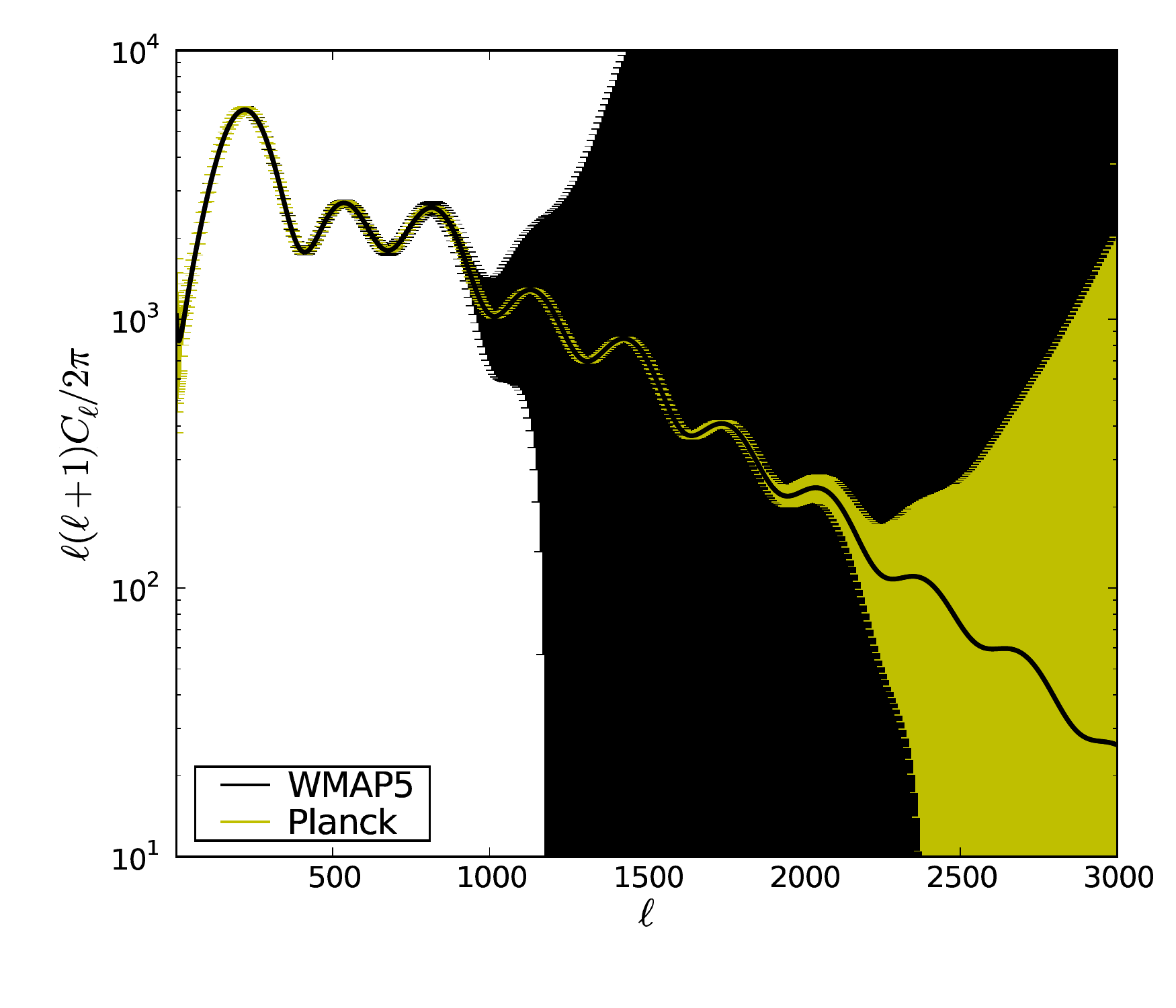}
  \caption{\label{fig:wmap5planck}Error models for the CMB temperature power spectrum used in the Fisher matrix predictions for WMAP5 (black) and Planck (yellow).}
\end{figure}

\begin{figure}[ht]
  \centering
  \plotone{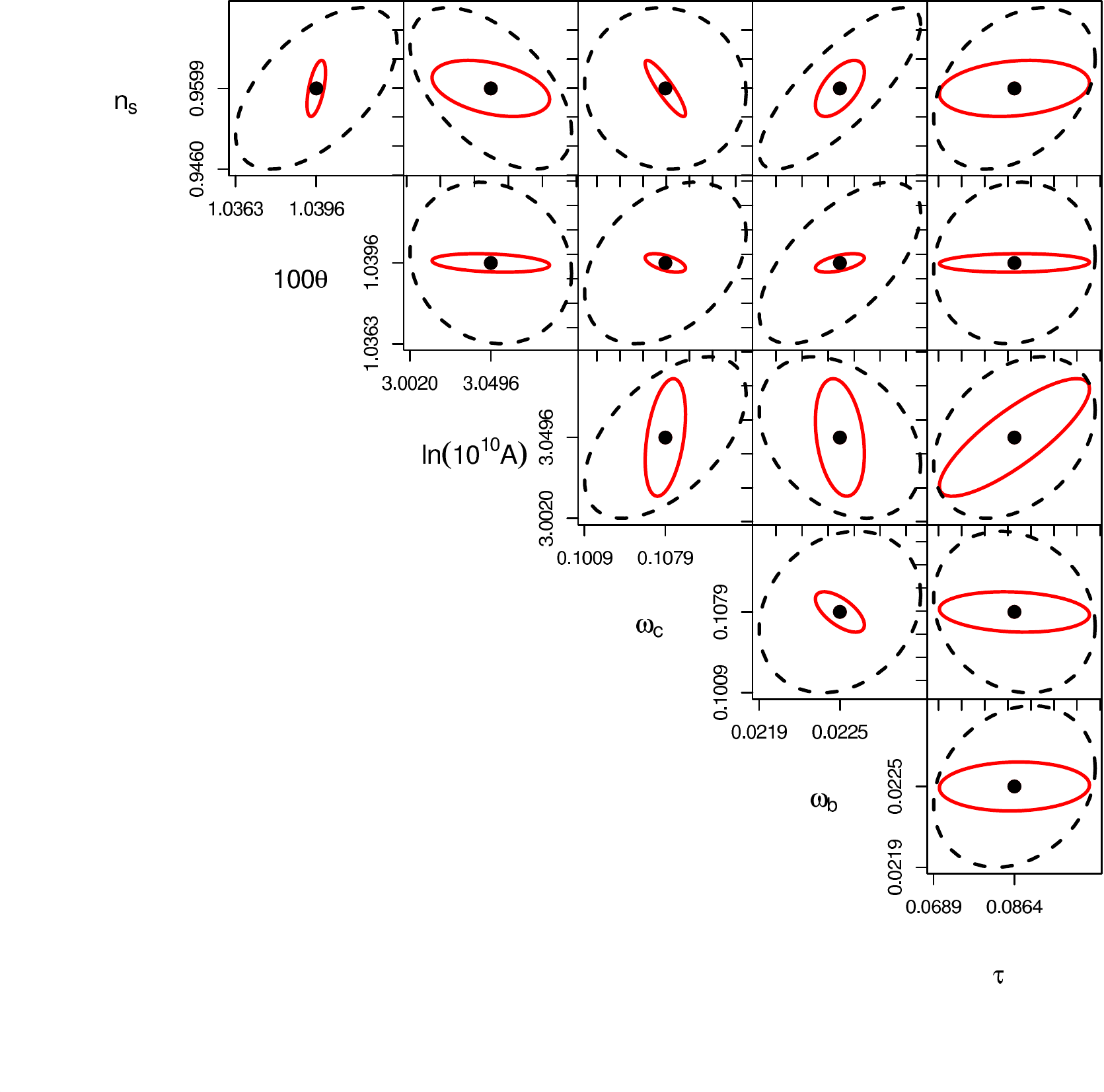}
  \caption{\label{fig:fishmats} 2-D marginal parameter constraints
    predicted by the Fisher matrices for Planck (red, solid) and WMAP5
    (black, dashed) noise and beam models.}
\end{figure}

We further include in our Fisher matrix a prior on a combination of optical depth and
primordial power spectrum amplitude similar to what is provided by WMAP5
polarization data.  Specifically, assuming the quantity $A e^{-2\tau}$
has negligible error (since this combination is determined very well 
from the temperature power spectrum),
we impose a Gaussian prior with width: 
\begin{equation}
  \frac{\sigma \left(A \tau^2\right) }{A \tau^2} \approx
  \frac{\sigma \left(e^{2\tau} \tau^2\right) }{e^{2\tau}\tau^2} \approx
  \frac{2(\tau+1)\sigma(\tau)}{\tau},
\end{equation}
with $\sigma(\tau) = 0.017$ from WMAP5\footnote{\url{http://lambda.gsfc.nasa.gov/product/map/dr3/params/lcdm_sz_lens_wmap5.cfm}}, which comes from the combination
of polarization data with temperature data and is the main contribution of
the polarization data to parameter constraints.  
In Figure~\ref{fig:wmap5planck} we show the size of
the power spectrum error bars for our WMAP5-like observations, and
those for our Planck-like observations.  In Figure~\ref{fig:fishmats}
we show the corresponding two-dimensional Fisher matrix contours.  

We were sufficiently satisfied with the accuracy and simplicity of the
emulator we developed, that we are releasing a version of it as a
publicly available code, called \textsc{Emu CMB}, described in
Appendix~\ref{sec:emuCMB}.  
There is another fast and accurate
code available \citep[\textsc{PICO}][]{pico,pico2} that has some advantages.
Most notably, \textsc{PICO} is a more complete emulator that provides
predictions for polarization as well as temperature power spectra and
includes models with non-zero spatial curvature.
\textsc{Emu CMB}'s advantages
are (1) its significantly simpler algorithm that has made it very easy
for us to make available implementations in multiple languages with no
need for external libraries, (2) the simple training algorithm can be
reimplemented quickly (using our public code if desired) for revised
CMB models, 
and (3) it extends to higher maximum
multipole moment ($\ell_{\rm max} = 5000$).  The latter quality is
especially important for analysis of high-resolution data, such as
from the Atacama Cosmology Telescope or the South Pole Telescope (SPT).

The design runs for this paper were all calculated using 
\textsc{CAMB}~\citep[2008 June release][]{camb} with accuracy settings {\tt
  accuracy\_boost = 4}, {\tt l\_accuracy\_boost = 4}, \\{\tt
  l\_sample\_boost = 8}, and without including gravitational
lensing\footnote{The effects of gravitational lensing are 
included in \textsc{Emu CMB}, since it is an important effect to include for
analysis of real high-resolution data.}. 
With this accuracy setting, numerical integration errors are
negligibly small.
We modified \textsc{CAMB} to extract the power spectra only at multipoles where
they are actually computed, skipping the default interpolation to all
multipoles.  For a maximum multipole of $\ell=3000$, this gives 448
power spectrum values using the default multipole spacing in \textsc{CAMB}. 

\subsection{Simulation design construction} 
\label{sub:simulation_design_construction}

Here we review two design procedures:  that used for CC and
that used for PICO.  We also introduce our new design procedure.  We compare
results from these three design procedures in Section~\ref{sec:emulator_results}.  

The CC design approach is ``Orthogonal Array Latin Hypercube Sampling'', or
OALHS~\citep{tang93,morris95,leary03,Rlhs}.  It begins with defining
the boundaries of a hypercube in a scaled parameter space -- all
parameter dimensions rescaled to fit in an interval between 0 and 1.
A grid in this hypercube is then defined.  In a two-dimensional
parameter space 
the $n_d$ design points are assigned to the grid in such a way that
each design point is the only occupant of its row and the only 
occupant of its column.  In a $3$-dimensional space, the occupation of
site $\{i,j,k\}$ means that sites $\{l,j,k\}$ are empty for all $l \ne
i$, sites $\{i,l,k\}$ are empty for all $l \ne j$ and sites $\{i,j,l\}$
are empty for all $l \ne k$.  The generalization to
$n$-dimensional spaces is straightforward.  The final step is to then set the
exact location of each design point within its grid cell.  This is
done using various criteria that generally aim 
to maximize the distance between points in {\em any}
lower-dimensional projection.  The fact that the points are
well spread out, in many different ways of quantifying ``spread out'',
improves interpolation to non-design points.   

The design for PICO is constructed from samples from a prior probability distribution
for the parameters and hence we call this design procedure PS for
``Prior Sampling''.  In particular, the PICO design is a draw of samples
from a converged MCMC run. The design is motivated
by the fact that the samples are drawn from exactly the region of the
parameter space where we desire accurate emulation
(i.e., the region with non-negligible prior probability). 

\begin{figure}[ht]
  \centering
  \plotone{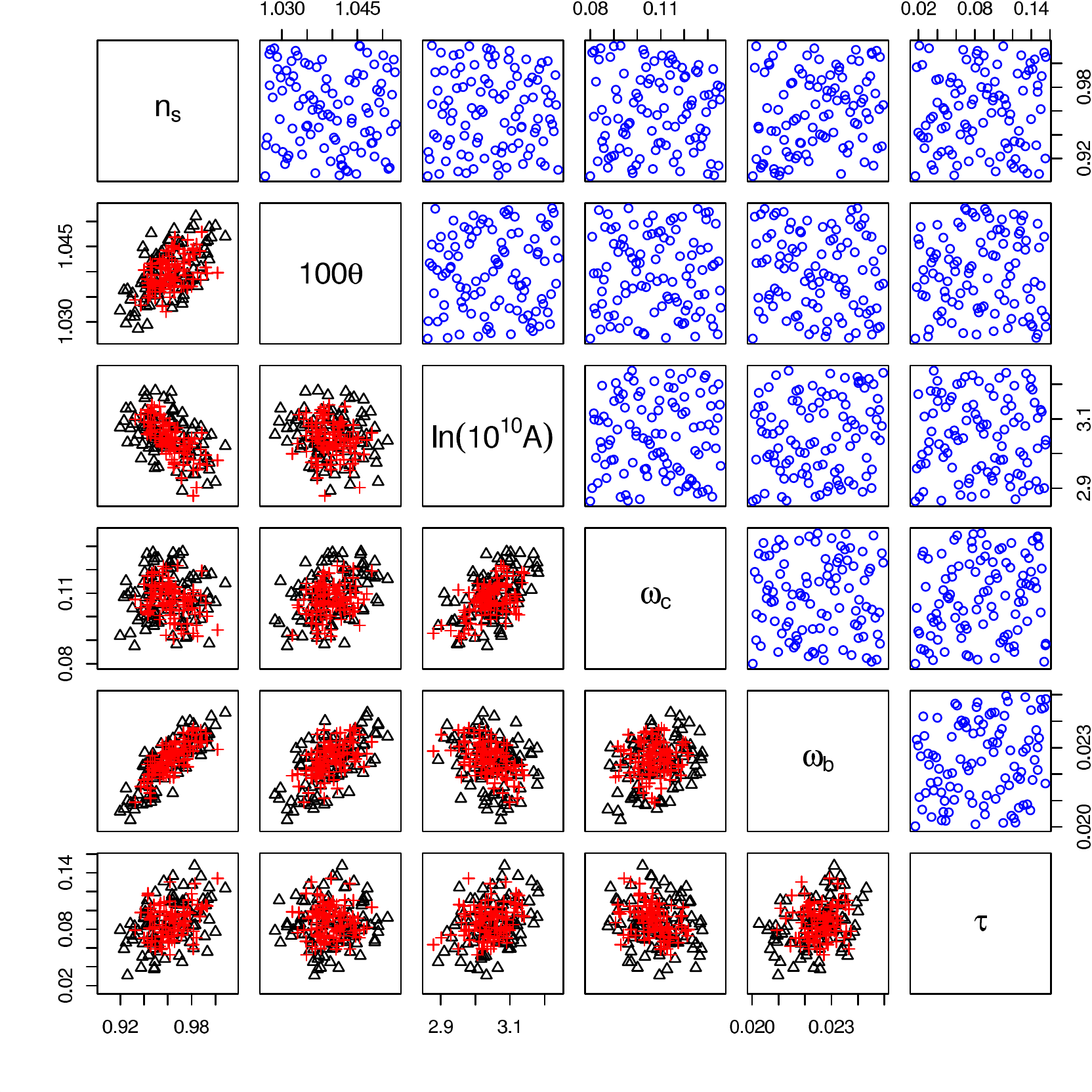}
  \caption{\label{fg:designprojections} Location of 100 simulation design runs in the 6-dimensional parameter space.  Upper triangle is for the OALHS design and the lower triangle is for the OALHSFS design (black triangles) and the PS design (red crosses).}
\end{figure}

The parameter bounds used in our OALHS design are given in
Table~\ref{tb:parambounds}.  The bounds have a width of eight times the
marginal Fisher matrix errors for the WMAP5 noise model, centered on
the fiducial point given in Column 2 of Table~\ref{tb:fidparams}.
Given these bounds and the number of design points, $n_d$, we draw a
realization of an OALHS using the \texttt{maximinLHS} function of the
R \textit{lhs} package~\citep{Rlhs}, which maximizes the minimum
distance between the design points within the Latin hypercube. 
\begin{deluxetable}{ccc}
  \tablewidth{0pt}
  \tablecaption{\label{tb:parambounds}Parameter range for OALHS
    design}
  \tablehead{
    \colhead{Parameter} & \colhead{Minimum} & \colhead{Maximum}
  }
  \startdata
  $n_s$                       & 0.9042 & 1.0156 \\
  $100\theta^*$               & 1.0264 & 1.0529 \\
  $\ln \left(10^{10}A\right)$ & 2.8592 & 3.2340 \\
  $\omega_c$                  & 0.07996 & 0.1359 \\
  $\omega_b$                  & 0.02000 & 0.02498 \\
  $\tau$                      & 0.01633 & 0.15653
  \enddata
\end{deluxetable}
All two-dimensional projections of a 100-point OALHS design are shown
in the upper triangle of panels in Figure~\ref{fg:designprojections}. 

Both the OALHS and PS design procedures have their advantages.  The
OALHS design fills the parameter space in an efficient way, providing
well-spaced anchor points for interpolation.  The PS design, on the
other hand, concentrates design points only in the region of parameter
space where we are actually interested in evaluating the emulator.  We
now present a design procedure that captures the advantages of both.   

If $p_{\theta}$ is the number of input parameters to our simulation,
we can compute a Fisher information matrix (based on the data error
distribution determining the parameter prior) from the summary statistic
of the simulations we wish to emulate using $2 p_{\theta}+1$
additional simulation runs; that is via $p_{\theta}$ two-point
numerical derivatives of the summary statistic and one fiducial point
used to calculate the summary statistic covariance model.  We use this
Fisher matrix to rotate the input parameter axes to a less-correlated
set of input parameters.  We then build an OALHS design in this
de-correlated parameter space, but reject all samples of the OALHS
design that are outside of a hypersphere of constant probability (at
quadratic order) centered on the fiducial parameter location.  We call
this design OALHSFS for OALHS samples inside a ``Fisher sphere''.  We
also use this decorrelated space for performing the interpolation of
the simulation design runs. 

The two-dimensional projections of a 100-point OALHSFS design constructed from a Fisher matrix using the WMAP5 noise model are shown as the black triangles in the lower panels of Figure~\ref{fg:designprojections}.  The design points are located within the hypersphere of radius 4 in the decorrelated parameter space (note that each parameter has unit variance in this space).  
We chose a radius of 4-$\sigma$ because this encloses over 98\% of the volume of a six-dimensional Gaussian distribution, which we judged to be sufficient for our emulator.  (Note that this is covering 98\% of the volume of the {\it prior} distribution.  The fraction of the posterior distribution covered will be much larger .)  The points of a 100-point PS design selected as random samples from an MCMC chain using the CMB power spectrum with the same WMAP5 noise model are overplotted as red crosses for comparison.  It is readily apparent that the OALHSFS design points are confined to the region of high prior probability as defined by the MCMC samples.

By choosing design points inside a hypersphere in a de-correlated
parameter space, we can achieve large reductions in the volume of the
space that must be covered by the design points as the dimensionality
of the parameter space increases.  In three dimensions, the volume of a
sphere embedded in a cube, such that the sphere's diameter is equal to
the length of a side of the cube, is nearly half the volume of the
cube.  In six dimensions, the hypersphere has a volume nearly 12
times smaller than that of the surrounding hypercube.  In principle then, if the interpolation error in our
emulator is largely determined by the mean distance between design
points, we should be able to achieve similar interpolation precision
using 12 times fewer design points in an OALHSFS design as in an OALHS
design in a six dimensional parameter space.  Or, for a fixed number of
design points, the OALHSFS design should have smaller interpolation
errors than the OALHS design.   

As shown in Figure~\ref{fg:designprojections} the PS design gives
similar volume savings over the OALHS design as does the OALHSFS.  
However, the PS designs do not consistently cover the design space and
can lead to very large interpolation errors in regions where the
design points are sparse 

The Fisher matrix approximates the shape of the likelihood about its
peak at quadratic order in the cosmological parameters.  Our OALHSFS
design will fail to encompass a region of constant probability if
higher order corrections to the shape of the likelihood are
significant.  The PS design does not suffer from this problem.  In
practice, it will be possible to measure the accuracy of the OALHSFS
probability contours by comparing with a PS design as we have done in
the lower panels of Figure~\ref{fg:designprojections}.  If necessary,
the OALHSFS design can be adjusted to allow for different ranges along
the different axes in the ``de-correlated'' parameter space.  That is
some axes might extend for, e.g., $\pm 4\sigma$ while some might
extend for $\pm 12\sigma$, etc. (where $\sigma$ is the marginal error
determined from the Fisher matrix).  

In addition to the prior parameter ranges, the parameter ranges along
each axis of the simulation
design could also be informed by the curvature of the mode amplitude
surfaces to be interpolated.  
In the case that a design axis has a 
range much shorter than the response surface curvature scale along a
given coordinate axis the design will provide an 
 estimate of the
curvature scale.  This could result in, e.g., the response surface
being modeled as less smooth than it really is. 
Information about the curvature of the response surfaces is contained in the
second derivatives of the power spectra with respect to the
cosmological parameters (while the Fisher matrix uses the first
derivatives of the power spectra).  The range of the design axes could then
be increased in directions where the curvature is very small (e.g., using
the mode amplitude with the smallest curvature to set the axis
length).
On the other hand, if
the response surface is known to have an exceptionally high curvature
along a given axis, the density of design points could be increased to
accurately sample the high curvature.
Because the emulators for our case study and \textsc{Emu CMB} work to an
acceptable precision we have not explored these ideas further.

It is always possible that when inferring parameter constraints from a data set via MCMC, the MCMC chain will move some steps out of the region of parameter space covered by the emulator design.  If the MCMC chain spends very little time in the region outside the design, it may possible to fall back on evaluating the actual Boltzmann code for analyzing the CMB.  For applications with more expensive simulations (such as $N$-body simulations) the emulator's can predictions can be extrapolated to get an initial MCMC chain that can then be used to inform the construction of a new design.

\subsection{Dimensional Reduction}
\label{sub:dimreduction}

To reduce the dimensionality of the function to be interpolated, both
the CC and PICO emulators perform a principle component decomposition
of the $n_y \times n_d$ matrix of simulation design runs $y_{\ell}^i\equiv
\ell(\ell+1)/(2\pi)\,C_{\ell}^{i}$ where $i=1,n_d$ and $\ell$ takes
$n_y$ values in the range $\left(2,\ell_{\rm max}\right)$.  The principal
components that contribute least to the variance over the design
are discarded.  We have found that keeping the 20
most-important modes is more than sufficient for our purposes, i.e.,
the errors made from neglecting the other ${\rm min}\left(n_y,n_d\right)$ modes are insignificant.  We quantify the errors from the truncated mode decomposition in two ways.  First, we have verified that using 20 PC modes allows us to reconstruct the power spectra at all design points in several OALHS designs to within 0.1\%.  Second, we used the Fisher matrix to estimate the cosmological parameter biases that would be induced from the systematic errors in the reconstructed power spectra~\citep[see][for details on the Fisher matrix methods]{huterer05b}.  For $p_{\mu} = 20$, the distribution of parameter biases had first and third quartiles $< 10$\% for all six of our cosmological parameters.  This means that in the case of perfect interpolation of the power spectra, the inferred parameter constraints using our emulator would have biases less than 10\% of the $1-\sigma$ marginal errors for both the OALHS and OALHSFS designs.

Following~\citet{heitmann06,habib07,schneider08} we first subtract the
($\ell$-dependent) mean of the design runs from each design power
spectrum and then scale the result by a single number so that the
combined entries of our $n_y \times n_d$ matrix of design runs have
variance one.  Denoting this centered and scaled matrix $\eta$, we
then perform a singular value decomposition, $\eta=\mathbf{UBV}^T$
where $\mathbf{U}$  
has dimension $n_{y}\times p$ ($p \equiv {\rm min}(n_{y},n_{d})$) with 
$\mathbf{U}^{T}\mathbf{U}=\ident_{p}$, $\mathbf{V}$ 
has dimension $n_{d}\times p$ with $\mathbf{V}^{T}\mathbf{V}=\ident_{p}$,
$\mathbf{V}\mathbf{V}^{T}=\ident_{n_{d}}$, and $\mathbf{B}$ ($p\times
p$) is a diagonal matrix of singular values.  Finally we define the
matrix of basis vectors, $\mathbf{\Phi}\equiv\frac{1}{\sqrt{n_{d}}}\mathbf{U}\mathbf{B}$
and $\boldmath{w}\equiv\sqrt{n_{d}}\mathbf{V}^{T}$ so that
$\frac{1}{n_{d}}\boldmath{w}^{T}\boldmath{w}=\ident_{n_{d}}$. 

Retaining only the $p_{\mu} \le p$ most significant columns of $\mathbf{\Phi}$ we
can write the scaled power spectrum as a sum over $p_{\mu}$ modes 
\begin{equation}
  \eta_{\ell}(\boldmath{\theta})  = \sum_{\mu=1}^{p_{\mu}} w_{\mu}(\boldmath{\theta}) \Phi_{\ell\mu}, 
\end{equation}
where $\Phi_{\ell\mu}$ is the entry of the matrix $\mathbf{\Phi}$ in row $\ell$
and column $\mu$, and $w_{\mu}(\boldmath{\theta})$ is the $\mu$th
(parameter-dependent) mode amplitude.

\begin{figure}[ht]
  \centering
  \plotone{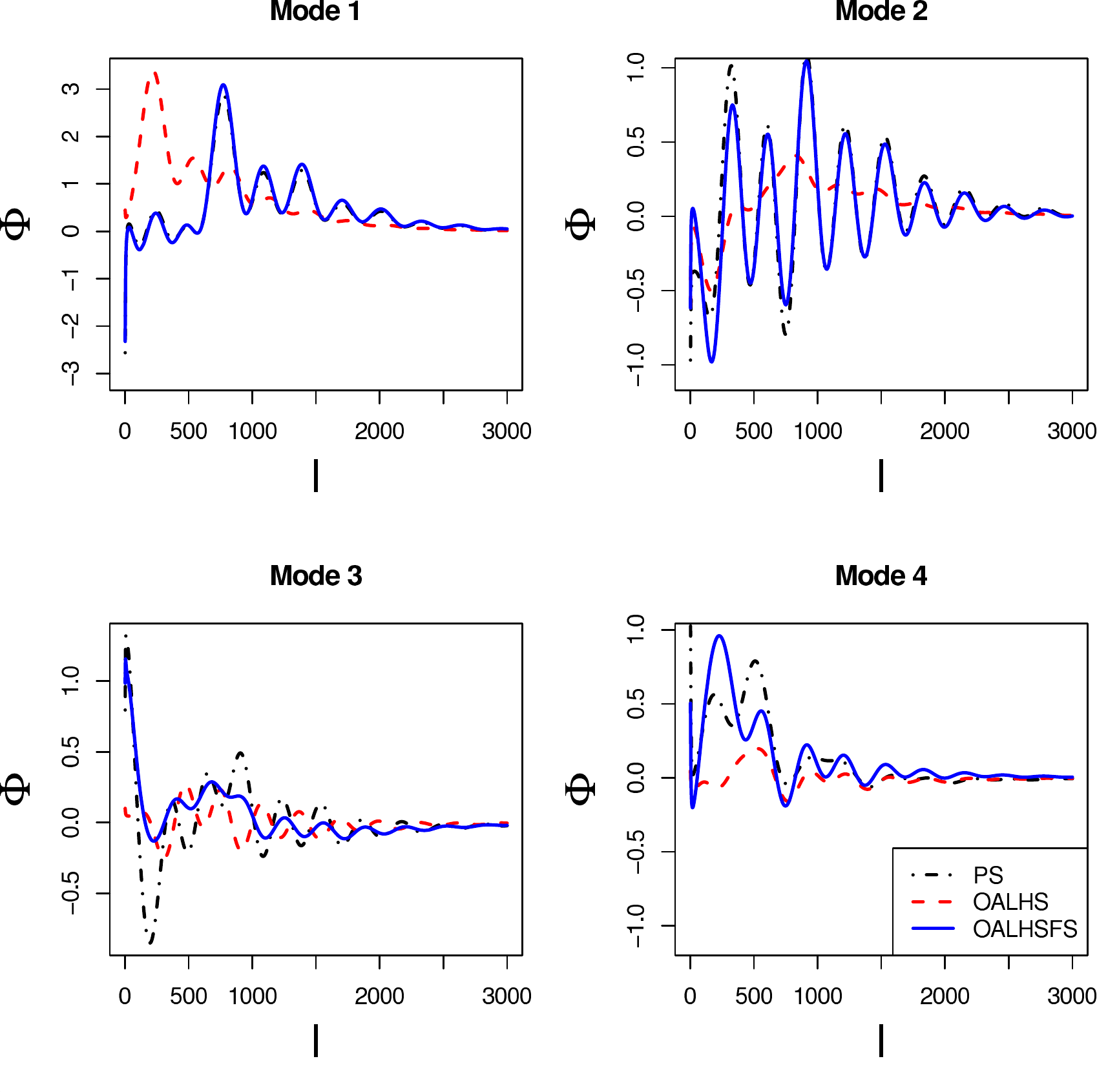}
  \caption{\label{fg:basisfunctions} The basis functions corresponding to the first four most significant principal components in the decomposition of our three types of 100-point designs.}
\end{figure}

In Figure~\ref{fg:basisfunctions} we show the first four PC 
basis functions for the power spectra in the three different
design types, PS, OALHS, and OALHSFS.  Because of the very different
volumes of parameter space covered, 
application of the PC algorithm to the design runs yields very different basis functions 
for the OALHS design versus the PS and OALHSFS designs.
This is a crucial difference because the basis functions must accurately
describe the design spectra at all multipoles in order to meet
the accuracy targets over the large dynamic range we are simulating.
Note that modes 1 and 2 derived from the PS and OALHSFS designs have several
oscillations, but are much smoother
when derived from the OALHS design.  This difference can be seen
later in the interpolation errors from the different designs
(Figure~\ref{fg:deltaCl}) where the OALHS design has larger, and more
oscillatory, errors at large multipoles indicating that the missing
oscillatory structure in the OALHS design basis functions leads to
missing structure in the reconstructed power spectra.
 In Appendix~\ref{sec:emuCMB}
we address the choice of basis functions again when we extend the
emulator for the CMB temperature power spectrum to $\ell=5000$.

\subsection{Interpolation}  
\label{sub:interpolation}

We compare two methods for interpolating the mode amplitudes
$w_{\mu}(\boldmath{\theta})$ between design points in the cosmological parameter
space.   

For the first interpolation method we again follow the CC framework
and model $w_{\mu}(\boldmath{\theta})$ as independent GPs for each
value of $\mu$.  We refer the reader to~\citet{habib07,heitmann09} for
details about the construction of the GP model for the mode
amplitudes.  In summary, for a fixed set of points
$\boldmath{\theta}_i$ ($i=1,\dots,n_d$), the
amplitudes $w_{\mu}(\boldmath{theta}_i)$ are modeled to have a multivariate Gaussian
distribution with mean zero and a parametrized covariance structure
that determines the smoothness of the interpolation.  The covariance
parameters are calibrated from a likelihood model conditioned on the
simulation design runs.  Because the GP framework infers distributions
for the covariance parameters rather than best fit values, the
error in the GP fit and interpolation can be propagated through to the
final cosmological parameter constraints.   

Our second interpolation method is to fit a $p_{\theta}$-dimensional
linear or quadratic polynomial to the design runs.  Compared to the GP
model, this has the advantage of being very fast to compute, but the
disadvantage of not propagating any interpolation error.  For small
numbers of design points, the order of the polynomial is limited (in
order to obtain a well-conditioned fit) and this can further limit the
accuracy of the polynomial interpolation.  We therefore expect GP
interpolation to outperform global polynomial fits when
the curvature of the mode amplitude surfaces (in any parameter
direction) exceeds the maximum polynomial order that can be fit with
$n_d$ design points.  A caveat is that GP models with our chosen covariance will also yield unacceptably large interpolation errors for extremely rapidly varying surfaces unless the number of design points is significantly increased.  So there is a
general limit on the smoothness of the mode amplitudes as functions of
cosmological parameters for such high-dimensional parameter spaces and limits on the practical number of design runs.

\begin{figure*}[ht]
 \plottwo{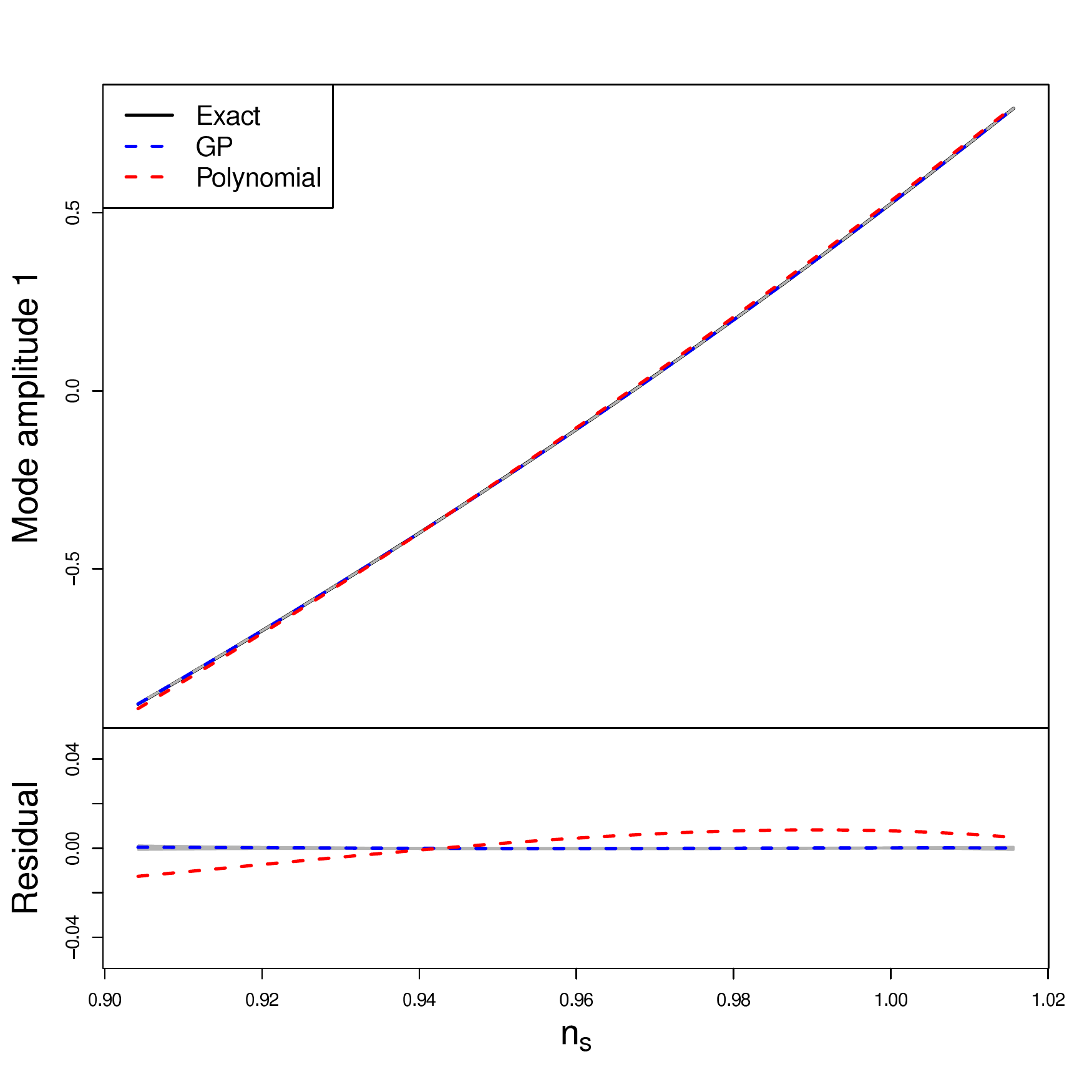}{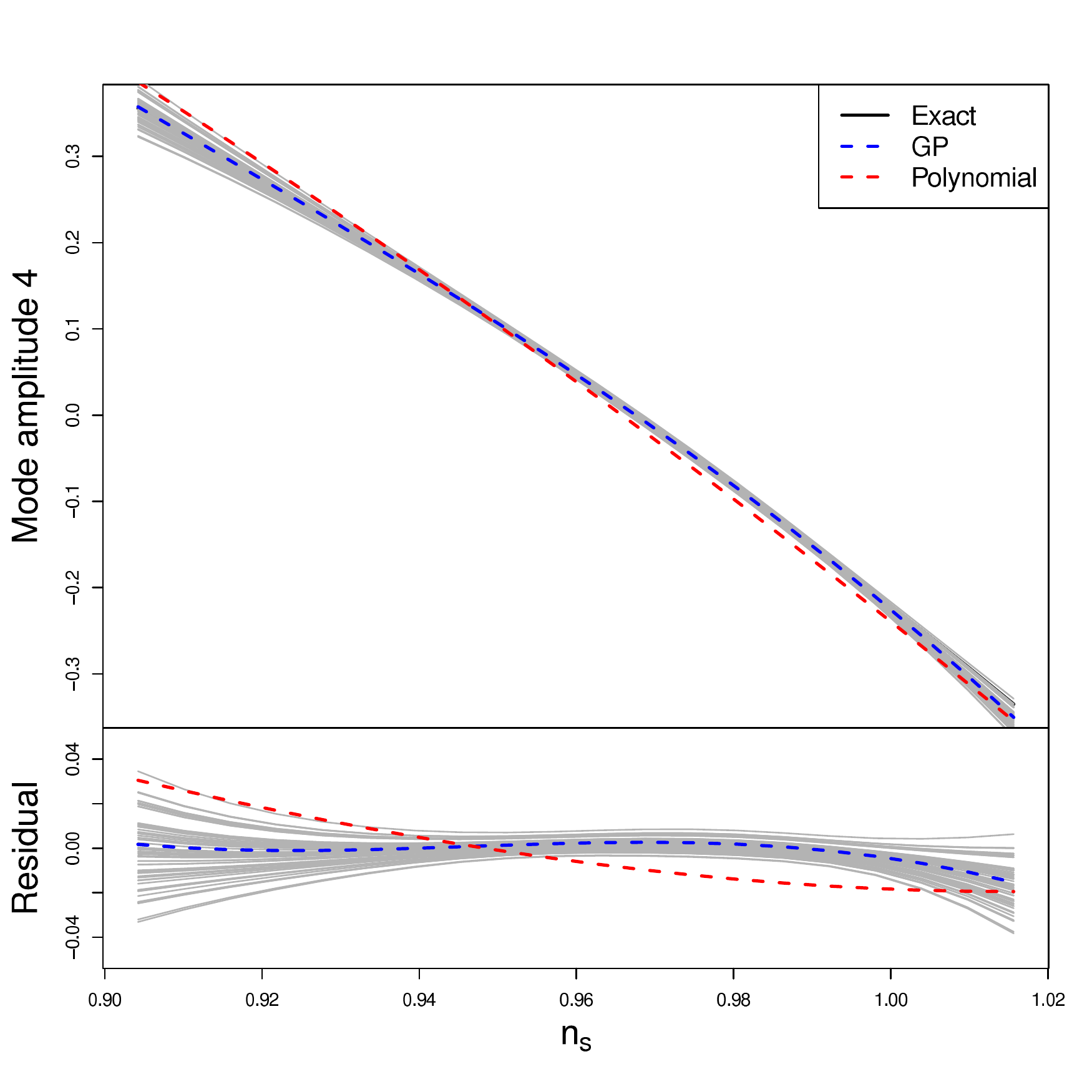}
  \caption{\label{fg:modesvsns}The first and fourth mode amplitudes
    for the OALHS 100-point design as a function of $n_s$, with all
    other comological parameters fixed at their central values.  The
    range of variation in $n_s$ is given by the marginal 4-$\sigma$
    errors as inferred by the WMAP5 Fisher matrix (i.e., the total
    width of the OALHS design along the $n_s$ axis). The gray lines
    show GP draws conditioned on the design runs using the
    maximum-likelihood GP parameters.  The blue lines show the mean of
    the GP draws.  The red lines show a polynomial fit that is quadratic order in the 6
    cosmological parameters.} 
\end{figure*}

Figure~\ref{fg:modesvsns} shows the first and fourth mode amplitudes
in the PC decomposition of the power spectra in the OALHS 100-point
design as a function of $n_s$ with all other cosmological parameters
fixed at their central values.  The range of $n_s$ covers the marginal
4-$\sigma$ error as derived from the Fisher matrix with the WMAP5
noise model.  The first mode amplitude in this case is a very smooth
function of $n_s$, while 
the less significant modes have a higher degree of curvature.  
(Note that the size of the residuals in the
mode amplitudes does not obviously translate into power spectrum
errors because of the centering and scaling of the power spectra done
before performing the mode decomposition.) 

\begin{figure}
  \centering
  \plotone{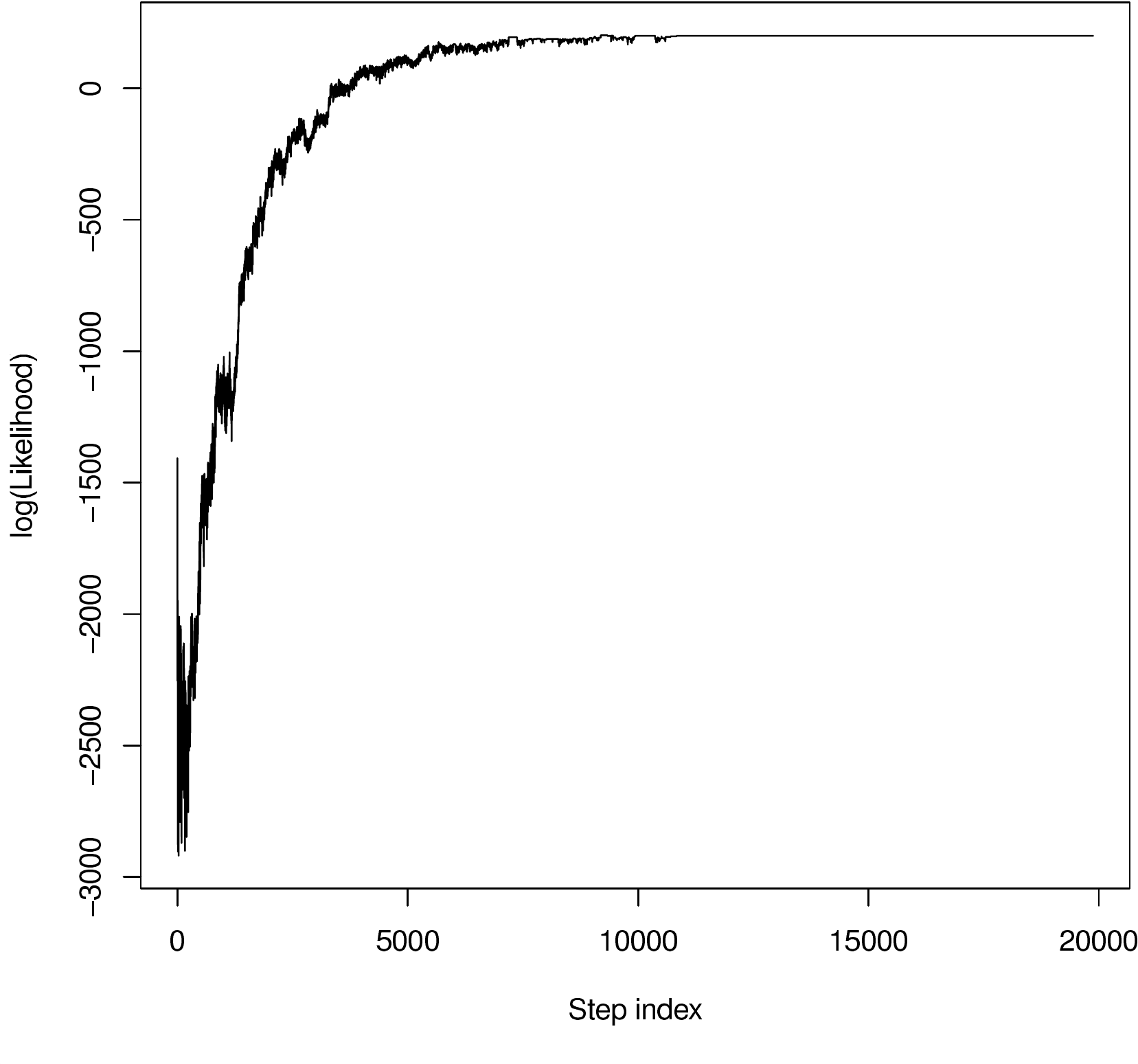}
  \caption{\label{fg:SAdiagnostics}Trace plot of the likelihood in an
    MCMC chain using simulated annealing to calibrate the parameters
    of a GP model for design interpolation.  This particular chain was
    calibrating a PS design.} 
\end{figure}

For our case study we have found the step of calibrating the GP
parameters to have intensive computatational and human-labor requirements.  With the GP covariance
parametrization used in the CC framework there are $p_{\mu}$ variance
parameters and $p_{\mu}\times p_{\theta}$ correlation parameters.  To
calibrate the GP model we must then find the maximum of the likelihood
in this high-dimensional parameter space.  We used MCMC to find the
vicinity of the GP likelihood peak, but found the most difficult step
to be initializing the chains in regions of non-negligible
probability.  A simulated annealing algorithm with a constant cooling
schedule turned out to be quite effective in getting our GP
calibration chains started.  Because the calibration of the GP model
can be a large computational obstacle in constructing an emulator we
highly recommend the use of a simulated annealing or similar
algorithm.    We do not, however, know of any algorithm that can
replace the human labor of assessing the convergence of the multiple
MCMC chains required to optimize the GP model.  
We demonstrate the performance of our simulated annealing algorithm
for calibrating the GP interpolation parameters in
Figure~\ref{fg:SAdiagnostics}. 

We also found the choice of the proposal distribution for the GP
correlation parameters in the MCMC algorithm to be crucial to
efficient calibration of the GP model.  We used independent uniform
distributions for each correlation parameter, but initially allowed
the width of the uniform proposal to be adjusted according to the
variance in the previous 1000 steps in the chain.  Because this
breaks the detailed balance requirement for convergence of the MCMC
chain, we fix the widths of the proposal distribution after a preset
number of chain steps (of order $10^5$). 

Because the evaluation of the GP likelihood at each chain step
requires inversion of an $n_d\times n_d$ matrix, the GP interpolation
method becomes unpractical as $n_d$ becomes very large.  This is
unfortunate because the interpolation accuracy should improve as $n_d$
increases!  We therefore expect GP models to provide a preferable
interpolation method for small $n_d$ and polynomial interpolation to
be preferred for large $n_d$.  For our case study, we actually find
that polynomial interpolation provides smaller interpolation errors
for a wide range of $n_d$.  We demonstrate the performance of
  polynomial interpolation for a different case study matching that of
  \citet{heitmann09} in Appendix~\ref{sec:coyoteUniverse}.

Our C++ code for calibrating GP models for interpolation~\citep[built
with the Scythe Statistical library][]{scythestatlib} is available for
download at \url{http://www.emucmb.info}.

\section{Emulator Results Compared} \label{sec:emulator_results}

\subsection{Designs Compared}
\label{sub:designs_compared}

\begin{figure*}[ht]
  \centering
  \plotone{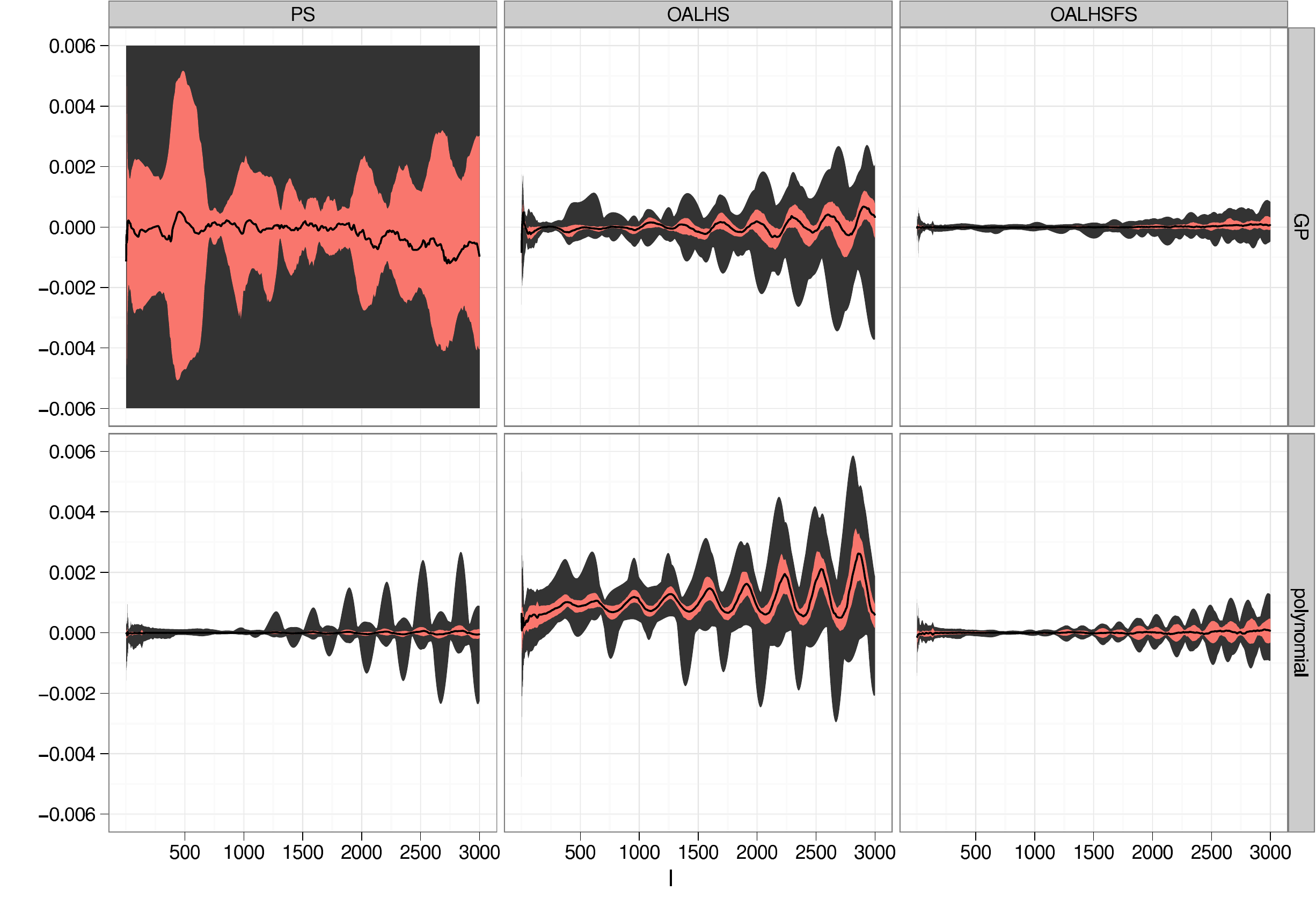}
  \caption{\label{fg:deltaCl} Fractional power spectrum interpolation
    errors using 3 different 100-point simulation designs: PS (left),
    OALHS (center), OALHSFS (right).  The interpolation is performed
    using GP (top) and quadratic polynomial (bottom) fits to the 20
    most significant principal components in each design, and is
    tested against 100 points in an additional PS design.  The dark grey
    bands show the minimum and maximum errors out of the 100 design
    points for each $l$ while the red (inner) band shows the errors within the
  first and third quartiles.  The central black line shows the median
  interpolation error.}
\end{figure*}

\begin{figure*}[ht]
 \centering
  \plotone{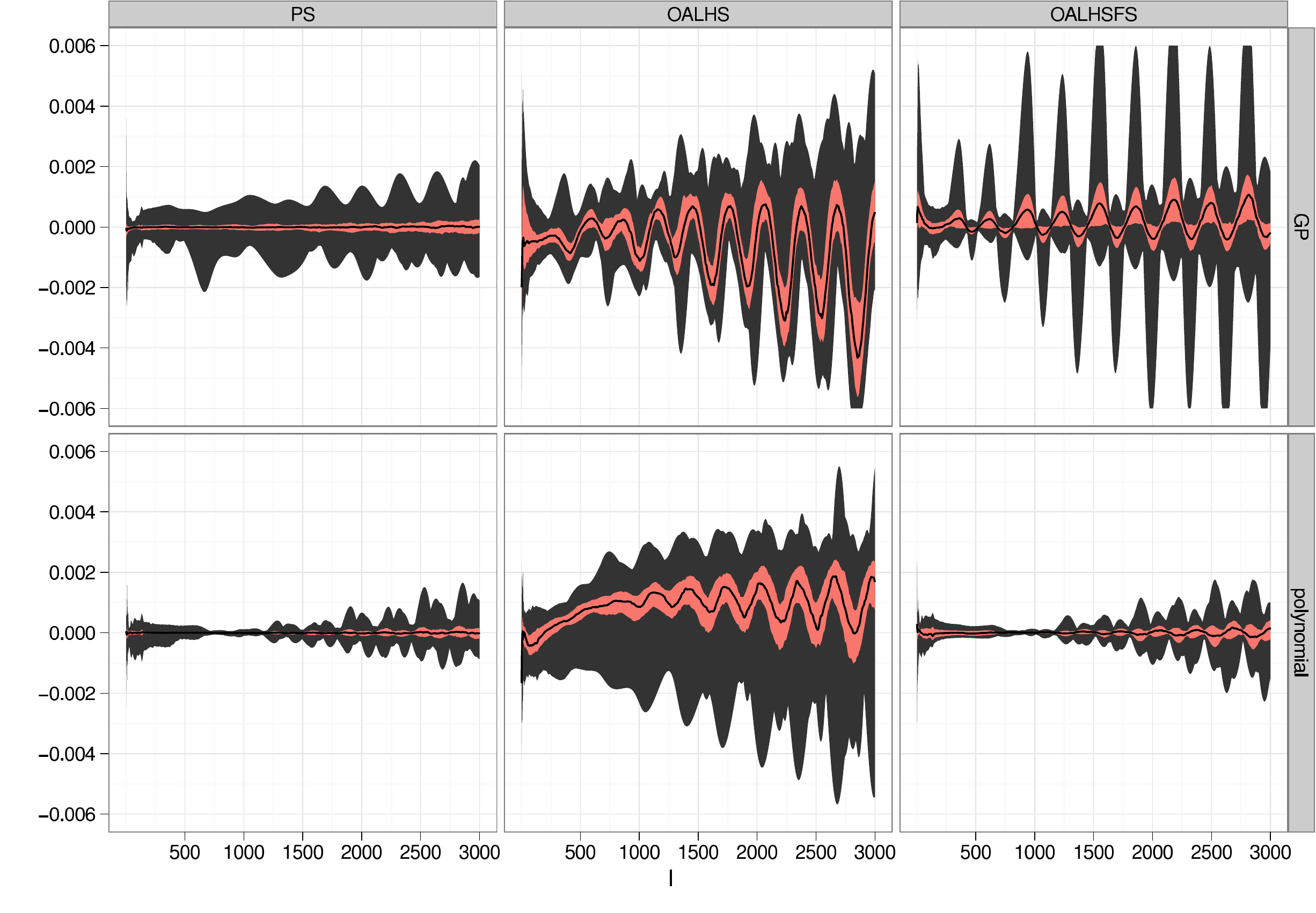}
  \caption{\label{fg:deltaClnd50} Same as Figure~\ref{fg:deltaCl} except for
  $n_d=50$.}
\end{figure*}

In Figure~\ref{fg:deltaCl} we show the fractional interpolation error for 100 test spectra using the three design
methods under consideration.  The test spectra were computed at points in our six-dimensional 
cosmological parameter space drawn from the posterior distribution for our WMAP5 noise model. 
Each interpolation method was trained on 100 design points.  (Note that we always show the 
expected value of the emulator; we  do not compute the random draws that are part of a complete GP
prediction.)  At these test points, the OALHSFS design has interpolation errors approximately two times 
smaller than the OALHS design at all multipoles and for both interpolation methods.  Using quadratic
polynomial interpolation, the PS design has comparable interpolation errors to the OALHSFS design,
but has very poor performance using GP interpolation.  
As we mentioned in Section~\ref{sub:simulation_design_construction}, the OALHSFS design is built in
a parameter-space volume that is $\sim1/12$ the size of the OALHS volume.  If the increased density
of simulation points is the dominant difference between the OALHSFS and OALHS designs, then
Figure~\ref{fg:deltaCl} gives some indication of how the interpolation accuracy improves with the 
increased density.  
However the interpretation is not straightforward as  the shape of the designs (spherical versus cubic)
may also be important.  That is, the design points outside the Fisher sphere in the OALHS design 
could, in principle, be helpful in reducing the interpolation accuracy, but we
have not pursued the separation of these effects further.  The main result that the OALHSFS design 
gives improved performance is sufficient for our purposes.
We show similar results, but for $n_d=50$ in
Figure~\ref{fg:deltaClnd50}.  The interpolation errors using quadratic
polynomials have similar relative performance between the three design
types as in Figure~\ref{fg:deltaCl}.  The GP interpolation results,
however, now show smaller errors using the PS design than using the OALHS
design.  The OALHSFS design using GP interpolation generally gives smaller errors than the
OALHS design, but there are a few test points where the OALHSFS design
gives worse performance for several multipoles.  
These results are consistent with our expectation 
that at a fixed number of design points $n_d$
the reduction in the mean distance
between design points in the OALHSFS and PS designs (see
Figure~\ref{fg:designprojections}) should improve the interpolation
accuracy.  Indeed we do see an improvement by a factor of $\sim 2$ in the
interpolation accuracy.   However, the poor performance of the PS
design with 100 design points and using GP interpolation demonstrates that the PS
design is a less reliable algorithm than OALHS-based designs 
for building an emulator using GP interpolation.  This could be
because the PS design leaves regions of parameter space very sparsely
covered compared to the OALHS designs (especially in higher
dimensions).  

\begin{figure}[ht]
  \centering
  \plotone{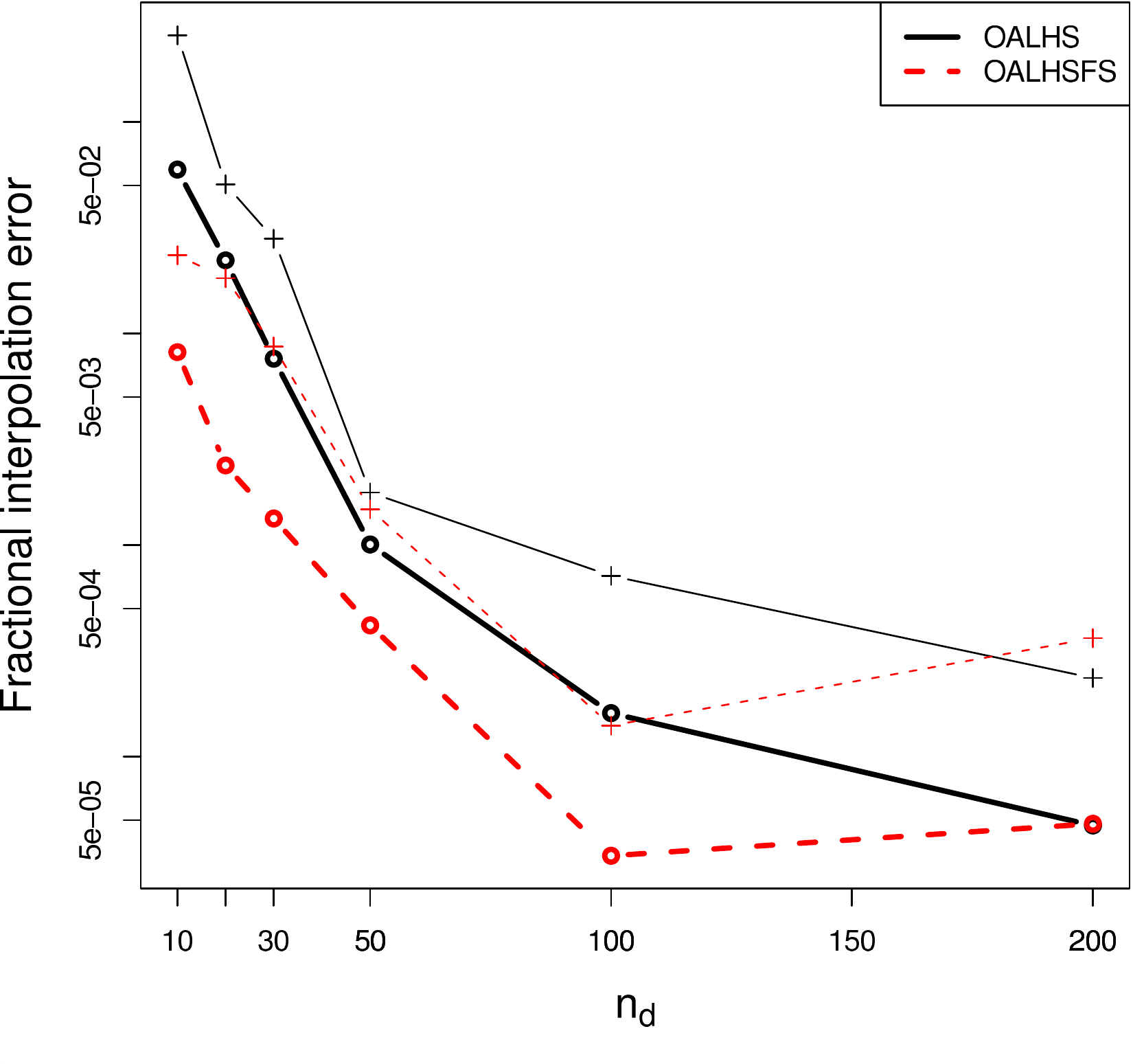}
  \caption{\label{fg:deltaClvsnd} Fractional error on the interpolated
    power spectra (using GP interpolation) averaged over $1500<\ell<2000$ as a function of the
    number of design points for 2 designs: OALHS (black -- solid) and OALHSFS
    (red -- dashed).  Thick lines show the maximum of the absolute value of the
    two quartiles of the error distribution over the 100 test points.
  Thin lines show the maximum absolute error.}
\end{figure}

Next we investigate how the accuracy gains of OALHSFS over OALHS at
fixed $n_d$ translate into a reduction in $n_d$ at fixed interpolation accuracy.  
In Figure~\ref{fg:deltaClvsnd} we show the quartiles and maxima of the
average fractional interpolation error in a band from $l=1500$ to
$l=2000$ (that is, we averaged the absolute
  value of the error over $l = 1500$ to 2000) versus $n_d$ for 100 test points
taken from a PS design.  We plot the
interpolation errors for both the OALHS and OALHSFS designs using GP
interpolation.  
Comparing the solid and dashed lines at fixed interpolation accuracy
(i.e., along a horizontal axis) we see the OALHSFS design requires
$\sim$25-70\% fewer design points than the OALHS design (with larger
savings at low $n_d$). 
One possible explanation for this trend is that some of the design
points in the OALHS design are not significantly contributing to the
reconstruction of the mode amplitude surfaces in the
(high-probability) region where the test points are actually located.
When $n_d$ is small, every design point should be important for
reducing the interpolation errors, so wasting even one point in the
OALHS design would degrade the emulator.
This demonstrates a main point of
the paper that the design space volume reduction offered by the
OALHSFS design is significant in minimizing $n_d$ to achieve a fixed
error tolerance.

\begin{figure}
  \centerline{
    \plotone{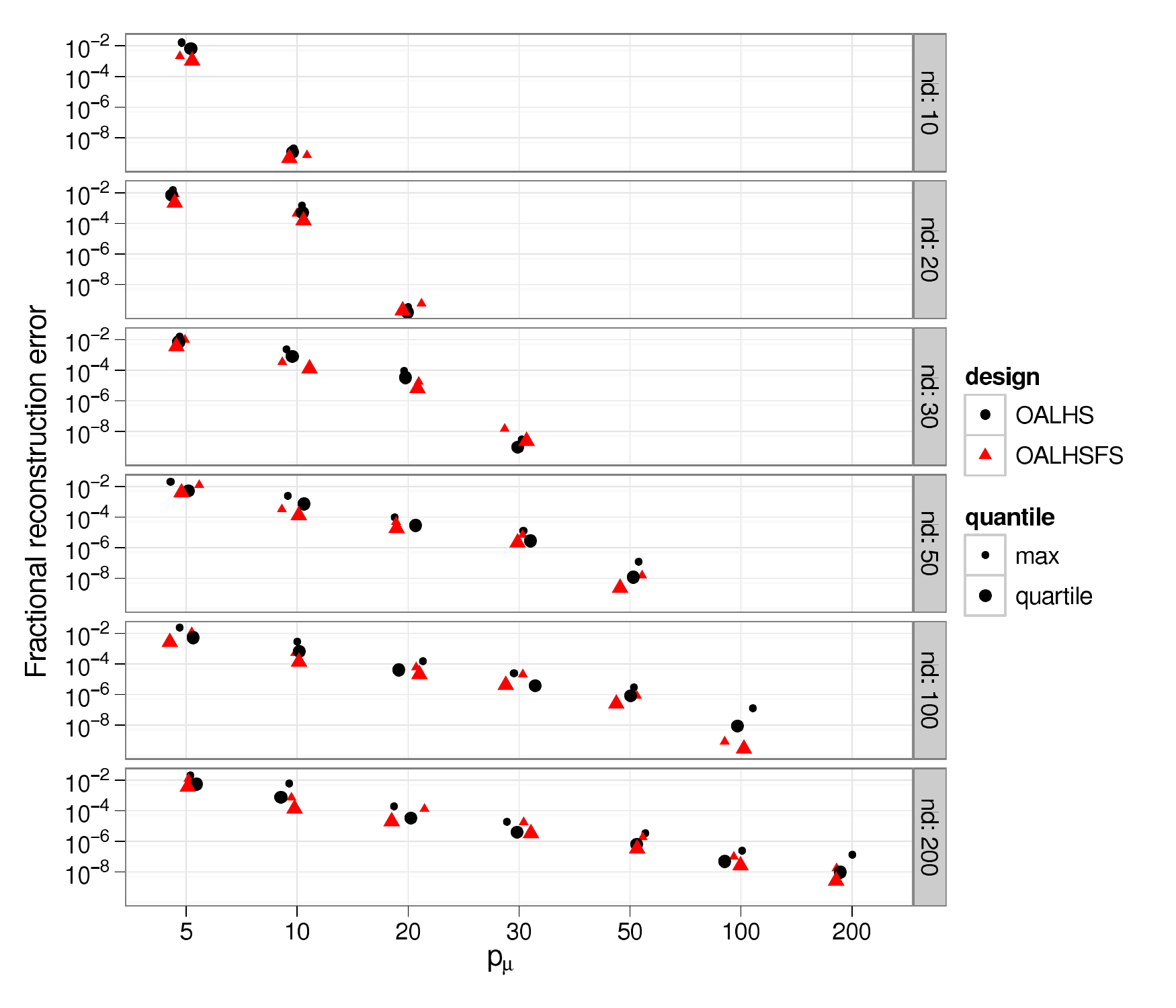}
  }
  \caption{\label{fg:deltaClvspmu} Fractional errors in the
     power spectra reconstructed from the truncated mode decomposition averaged over $1500<\ell<2000$ as a function of the number of Principal Components retained.  Our standard choice throughout the paper is $p_{\mu}=20$.  The power spectrum errors are computed at each design point  for each $n_{d}$ considered in Figure~\ref{fg:deltaClvsnd}.  (See the Figure~\ref{fg:deltaClvsnd} caption for the definitions of the quantile labels).  Note the points have been offset horizontally for clarity.}
\end{figure}

In order to understand when the errors in the truncated mode decomposition of the power spectra are significant relevant to the interpolation errors, we plot the fractional errors in the reconstructed power spectra as functions of $p_{\mu}$ at each design point ({\it i.e.} without any interpolation) in Figure~\ref{fg:deltaClvspmu}.  The reconstruction errors are rapidly decreasing functions of the number of retained PC modes, as expected.  For the choice used throughout this paper of $p_{\mu}=20$ for $n_{d} \ge 50$, the errors from the truncated mode decomposition are adequate for our error tolerance of $10^{-3}$ for both the OALHS and OALHSFS designs.  However, from Figure~\ref{fg:deltaClvsnd} we can see that with $p_{\mu}=20$ the mode reconstruction errors are non-negligible for the $n_{d}=100,200$ designs, which would explain why the total fractional power spectrum errors (mode reconstruction + interpolation) do not steadily decrease for $n_{d} > 100$.

\subsection{Interpolation Methods Compared}
\label{sub:interpolation_compared}

\begin{figure*}[ht]
  \centerline{
    \plottwo{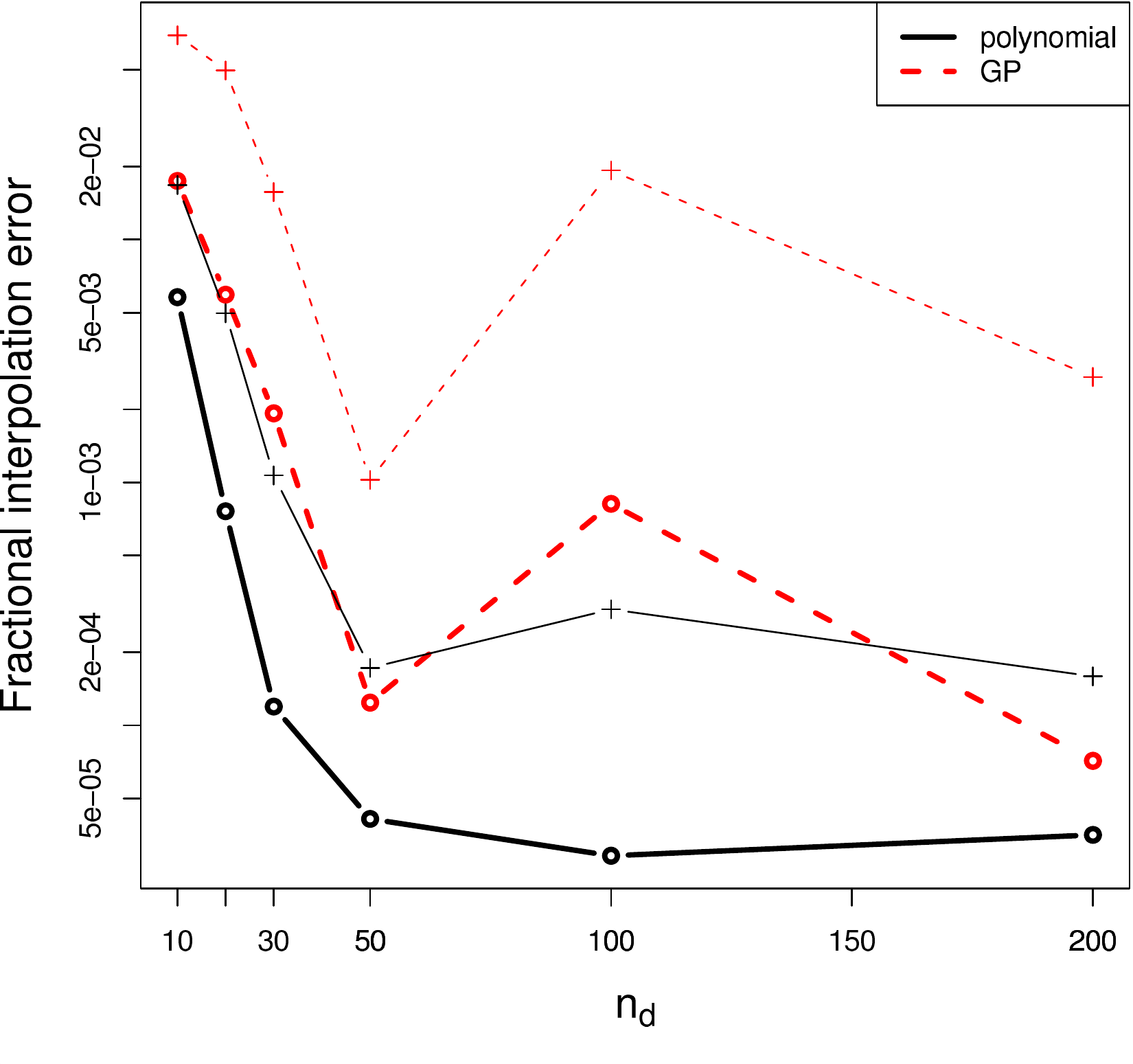}{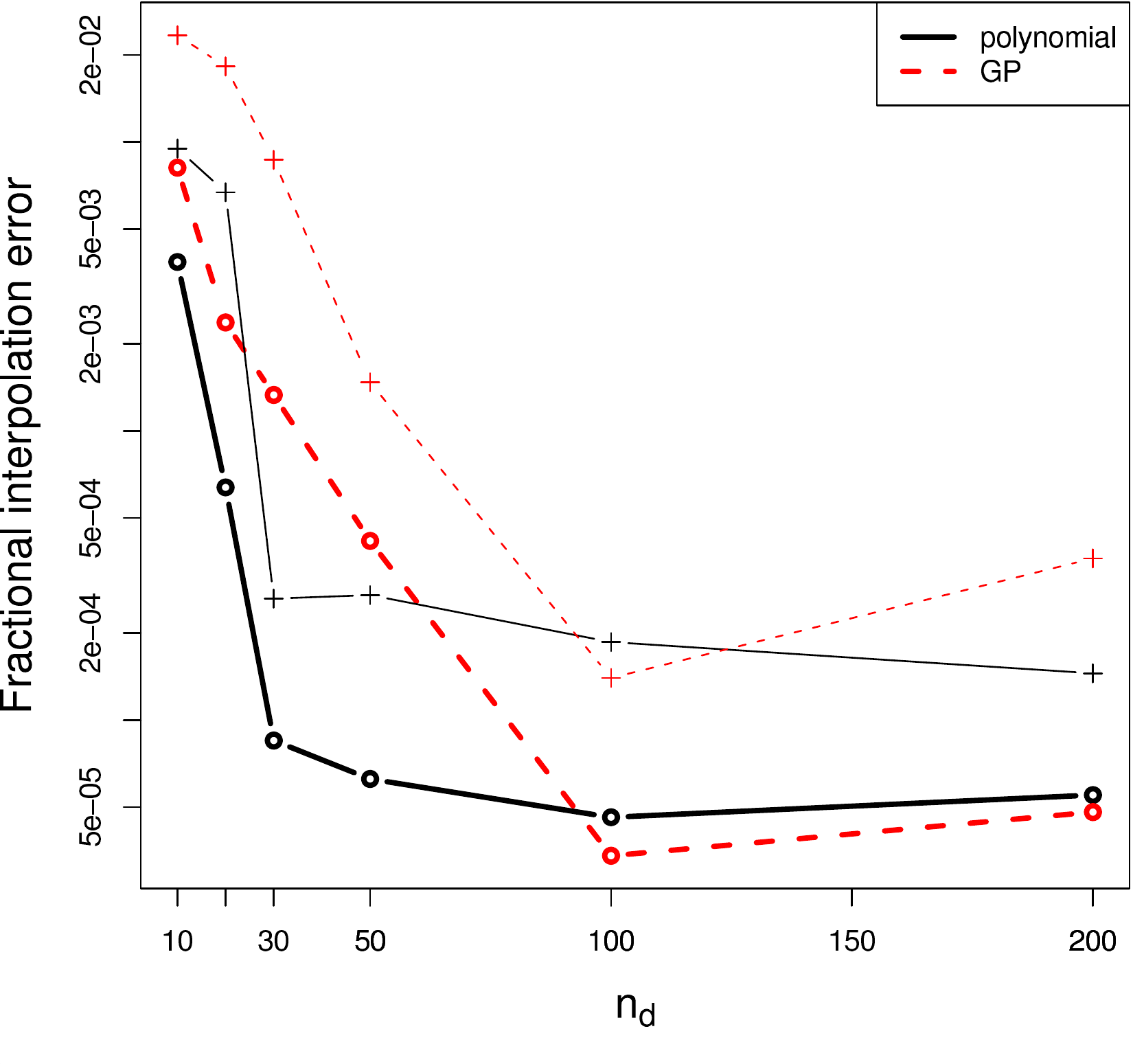}
  }
  \caption{\label{fg:deltaClinterpcomparison}Fractional error on the
    interpolated power spectra averaged over $1500<\ell<2000$ as a
    function of the number of design points using different interpolation
    methods: quadratic polynomial (black -- solid)
    and GP (red -- dashed).  The left panel shows the interpolation
    errors in the PS designs while the right panel shows the OALHSFS
    designs. Thick lines show the maximum of the absolute value of the
    two quartiles of the error distribution over the 100 test points.
  Thin lines show the maximum absolute error.}
\end{figure*}

In Figure~\ref{fg:deltaClinterpcomparison} we compare the mean interpolation
errors in the band $1500 < l < 2000$ using the GP model and a
quadratic polynomial fit to the
OALHSFS and PS designs as functions of $n_d$.  
In all of the designs
considered  polynomial interpolation provides smaller
interpolation errors than GP interpolation for small $n_d$.  For the
OALHSFS design, the interpolation methods give comparable errors for
large $n_d$.  
The PS designs have
sporadic error ranges, indicating the large variation in the covering
of the design space test points with different PS design realizations.

The (first or second order) polynomial interpolation performs so well because the response
surfaces of each mode amplitude are so smooth as functions of the six active
cosmological parameters.  The GP models also require smooth response
surfaces, but we expect the low-order polynomial fits to be worse than
the GP fits for a given correlation length in the response surface as
the correlation length is decreased.  This is because the covariance
structure in the GP model has more flexibility than a (global) second
order polynomial.  Therefore, for some applications the fast-to-compute 
polynomial interpolation will no longer suffice. The relative
performance of different interpolation methods will always be
problem-specific.  However, we explore
designs for the matter power spectrum in
Appendix~\ref{sec:coyoteUniverse} and show that the polynomial
interpolation is again comparable to the GP interpolation
demonstrated in~\citet{heitmann09} for similar designs for most of the
dynamic range in the simulated power spectra (except for the smallest
scales).  So, for both the CMB and matter power spectrum, an emulator
constructed using an OALHSFS design and quadratic polynomial
interpolation has similar performance to an emulator using an 
OALHS design and GP interpolation.
As the number of design points increases, both interpolation
methods should converge to arbitrary accuracy as higher-order polynomial fits
become well-conditioned.

\section{Conclusions}\label{sec:conclusions}
We have demonstrated an improved method for generating simulation
designs for interpolation based on selecting OALHS inside a
spherical (rather than cubical) volume in a whitened cosmological
parameter space defined by the Fisher matrix.  Our improved design
algorithm (OALHSFS) offers a significant improvement in design
efficiency over the OALHS algorithm; giving a factor of $\sim2$
reduction in the number of design points ($n_d$) required to reach a fixed
emulator accuracy.  We also compared the OALHS-based designs with
designs constructed by drawing samples from the prior distribution for
the cosmological parameters (which we call prior sampling or PS
designs).  Because both PS and OALHSFS designs concentrate design points in
the volume of parameter space with significant prior probability, they
both yield smaller interpolation errors for a fixed $n_d$ than 
OALHS designs when $n_d$ is large.  When $n_d$ is small, the PS
designs give poor coverage of the parameter space; giving erratic
interpolation errors depending on the particular PS design
realization.  This is not surprising because it is exactly the problem
that OALHS designs are meant to solve.  However, we take the superior
performance of both the PS and OALHSFS designs in the large $n_d$
limit ($n_d=100$ for our case study in six-dimensional parameter space),
as evidence that the OALHSFS design retains the advantages of the
OALHS design while also capturing the benefit of limiting design
points to regions of significant prior probability.

We compared the GP interpolation methods used in
previous literature of the emulator framework for cosmological
simulations with simple, global polynomial interpolation over the
design space.  We find that second order polynomial interpolation gives
interpolation errors that are only slightly worse than a GP model with
optimized correlation parameters.  This is a useful discovery because
the GP models can be computationally challenging to calibrate while a
polynomial fit via least-squares minimization is very fast.  For PS
designs we found it is not always possible to fit a GP model, while
second order polynomial interpolation still gives errors similar to those
using an OALHSFS design.  For applications where low-order polynomial
interpolation does not provide sufficient precision, we note that the
GP model calibration done via MCMC can be significantly hastened by
using a simulated annealing algorithm to search for regions of high
likelihood in the GP parameters.

Making use of the simplicity of polynomial interpolation, we are releasing code to generate
emulators of the CMB power spectra whose computational limitation is
only the time it takes to run the Boltzmann code to generate spectra
at the design points (available at \url{http://www.emucmb.info}).  We
demonstrate one such practical emulator in 
Appendix~\ref{sec:emuCMB} for the TT power spectrum reaching $\ell > 5000$ and including lensing
contributions.  This emulator uses 100 design points to achieve interpolation errors less
than 0.6\% in the power spectrum for all $\ell$.  
Python and IDL scripts as well as a module for
CosmoMC~\citep{cosmomc} for generating predicted power
spectra from this emulator are available at the same URL.  

Because of the similar performance of the two interpolation methods we
investigated, we conclude that the simulation design is the most
significant component of the emulator framework introduced in
\citet{heitmann06, habib07,heitmann09}.  The mathematically
consistent GP framework becomes important when it is necessary to
propagate the errors in the design interpolation or when optimizing
the interpolation for a given OALHS-based design.

As mentioned by other recent work~\citep{heitmann06, habib07,
  heitmann09} the simulation emulator technology becomes most
interesting for emulating computationally expensive simulations such
as those needed for predicting tracers of the matter power spectrum
(e.g., weak lensing or galaxy spectra).  In this case $N$-body
simulations of the dark matter distribution can take days or weeks to
run, which puts severe constraints on the number of design points that
can be considered.  In addition to the matter power spectrum, the need
for predicting the outputs of suites of simulations over cosmological
parameter space has recently been recognized for understanding the
deviation from universality of the halo mass
function~\citep{manera09}, the distribution of progenitor masses in
the conditional mass function~\citep{neistein09}, 
computing the 1-point probability distribution function of the
Ly$\alpha$ flux~\citep{viel09}, and mapping the input space of
semi-analytic models of galaxy formation~\citep{bower09a, bower10}.  We expect
the methods developed here to be useful for the first three 
applications.  However, the problem of mapping the input space for
simulations with a large number of parameters is qualitatively
different.  Our design based on the Fisher matrix assumes that the
prior probability distribution in the input parameter space is
unimodal and that the maximum of the prior probability distribution
is known a priori.
The success of the interpolation methods
discussed in this paper also assume that the ranges of input
parameters are sufficiently small that the outputs are (quite) smooth
functions of the input parameters.  Both of these assumptions are
broken when mapping the input space for semi-analytic galaxy formation
models so this should be considered a separate problem for emulator
construction \citep[see][for how OALHS designs are useful for such problems]{bower10}.  

Recently, \citet{angulo10} have shown that
statistics such as the power spectrum (as well as mass functions and
merger trees) can be obtained at different cosmological parameter
values using appropriate rescalings of a single $N$-body simulation.  If
this method proves to be successful in further tests, it could
alleviate the strict bounds on the number of design points that are
feasible for statistics derived from cosmological $N$-body simulations.
In this case, the advantages of GP interpolation methods would become less
useful (namely the ability to optimize the GP model to achieve minimal interpolation errors and to propagate the interpolation errors).  Alternatively, using the methods of \citet{angulo10} could
make it feasible to construct emulators of more sophisticated
statistics derived from light cones requiring many $N$-body simulations
for a single point in parameter space.  We plan to pursue this in
future work.

\acknowledgments
We thank David Higdon, Charles Nakhleh, Salman Habib,
  Katrin Heitmann, Ian Vernon, Richard Bower, and Shaun Cole for useful
  conversations.  We also thank Chad Fendt for reviewing
  our work and helping us understand how it relates to \textsc{PICO}
  and Alexander van Engelen and Gil Holder for providing their MCMC chain sampling from the
  WMAP5 posterior distribution.  We thank the anonymous referee for many useful suggestions to improve the manuscript.
This work was supported at UC Davis by NSF grant 0709498 and
an award from NASA's Jet Propulsion Laboratory.

\bibliography{gpm}

\appendix

\section{Emu CMB: a CMB TT power spectrum emulator for current
  experiments}\label{sec:emuCMB}
In this section we present a CMB TT power spectrum emulator suitable for
use in parameter estimation algorithms from
Planck\footnote{\url{http://www.esa.int/esaMI/Planck/index.html}}, ACT\footnote{\url{http://www.physics.princeton.edu/act}}, SPT\footnote{\url{http://pole.uchicago.edu}}, or similar
experiments.  We use \textsc{CAMB}~\citep{camb} to generate CMB TT power spectra with the
same six cosmological parameters as in the main body of the
paper\footnote{We revert to the more common definition of the
  primordial amplitude defined at the pivot scale $k=0.05 h$Mpc$^{-1}$.}, but out to
$\ell_{\rm max}=5238$ and including the lensing distortion of the
power spectrum.  We have not concerned ourselves with modeling
uncertainties such as those due to recombination.  Instead, we regard
this emulator as proof that a practical tool for current data analyses
can easily be constructed.  Emulators for revised models can be
quickly constructed using the same algorithm.  We provide scripts for
this purpose as well as details of the exact \textsc{CAMB} settings that
generated the results presented in this section at \url{http://www.emucmb.info}.

\subsection{Building the emulator}
We use an OALHSFS design with $n_d=100$ covering the WMAP5
4-$\sigma$ error bounds as determined by our Fisher matrix described
in Section~\ref{sub:cmbspectra}.  The design runs for \textsc{Emu CMB} use the
2009 October release of \textsc{CAMB} (different from the case study in the
main body of the paper) with the ``curved correlation function''
lensing implementation including nonlinear corrections calculated by
\textsc{HALOFIT}~\citep{challinor05}.  We again use accuracy settings {\tt accuracy\_boost = 4} and
{\tt l\_accuracy\_boost = 4}, but build the emulator from the power
spectra as output by \textsc{CAMB} at every multipole rather than extracting
only the uninterpolated power spectra values.  We set {\tt
  l\_sample\_boost = 4} to determine the number of multipole values
where the power spectra are actually computed.  See
\url{http://www.emucmb.info} for the complete \textsc{CAMB} parameter files.

\begin{figure*}
  \centerline{
    \plottwo{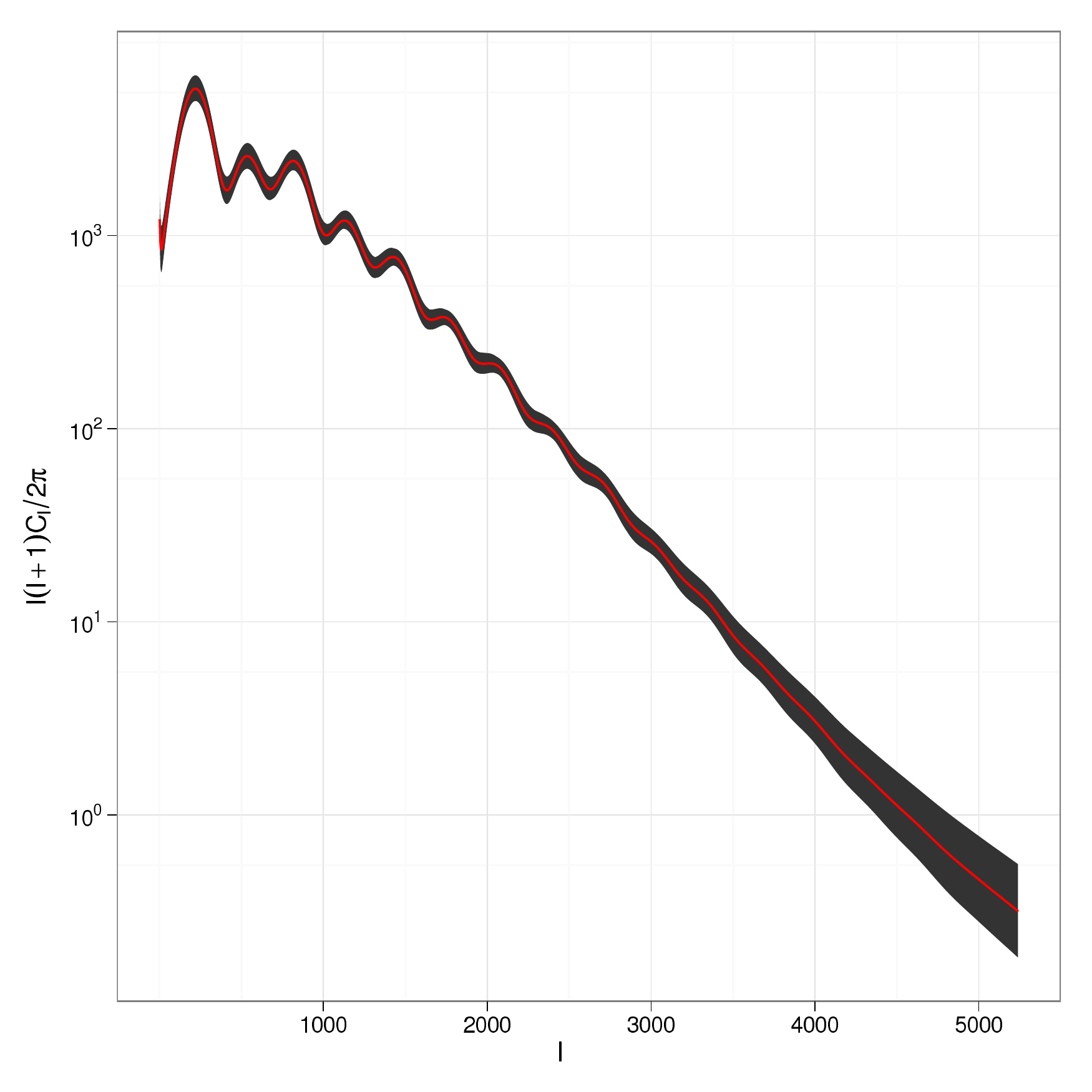}{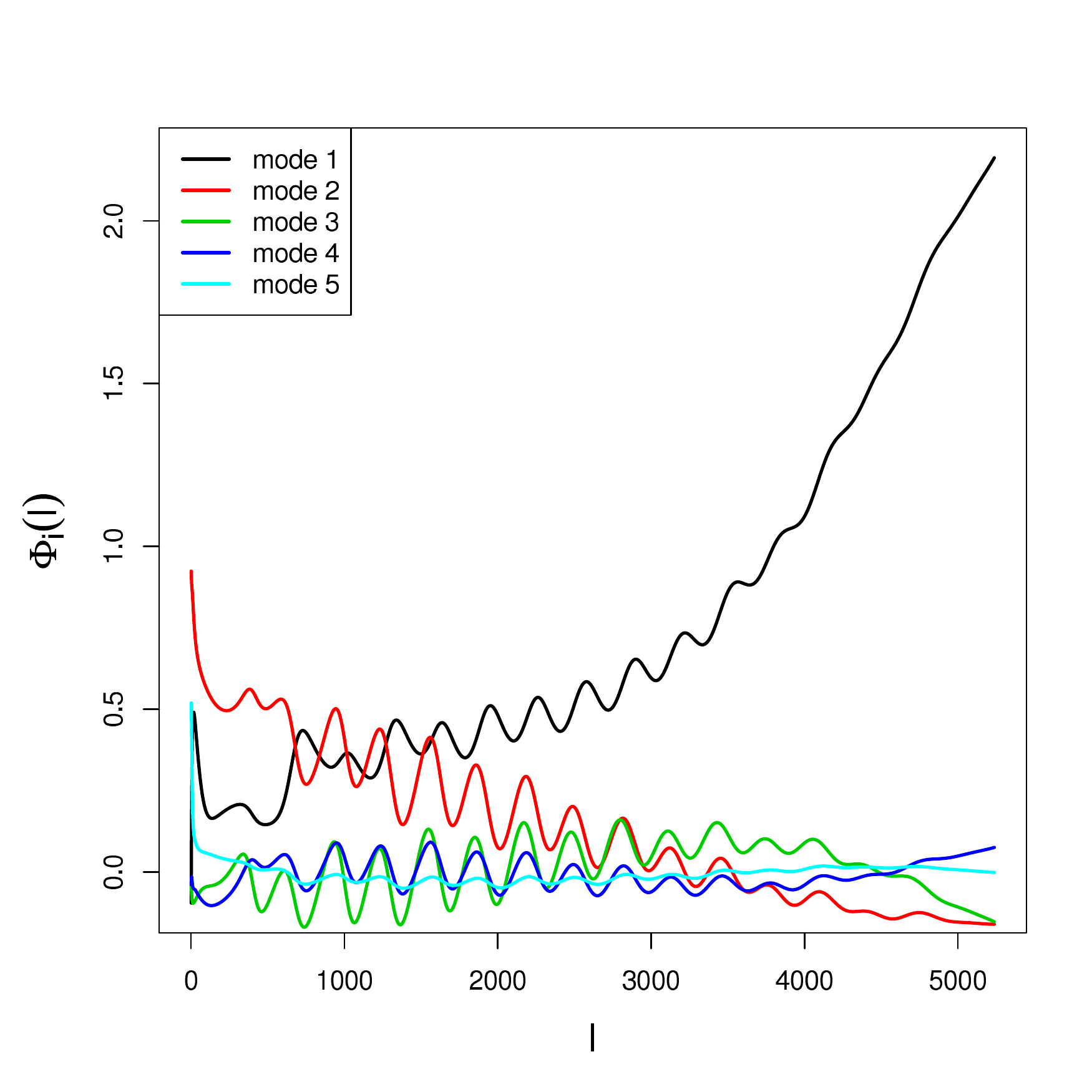}
  }
  \caption{\label{fg:emu_basis_functions} Left: TT power spectra at
    each of the 100 design points for \textsc{Emu CMB}.  The mean of the design
  runs is shown in red.  Right: basis functions for the 5 most
  significant principal components of the 100 design runs $\log(\ell(\ell+1)/(2\pi)\,C_{\ell})$.}
\end{figure*}

Also in contrast to the main body of the paper, we perform the
dimension reduction of the design runs on $y^i_{\ell}\equiv
\log\left(\ell(\ell+1)/(2\pi)\,C_{\ell}^i\right)$ ($i=1,n_d$).
Performing a PC decomposition on the logarithm of the power
spectra yields basis functions that better describe the lensing
variations at $\ell \gtrsim 3000$ across the design points as shown in
Figure~\ref{fg:emu_basis_functions}. 

As in the main text, we fit global second order polynomials to each of the first
20 mode amplitudes to perform the interpolation over the design
space.  Using only the first 20 PCs allows
reconstruction of the power spectra at the design points to less than 0.1\% accuracy for
$\ell \lesssim 100$ and to less than 0.01\% accuracy for $\ell \gtrsim 100$.

\subsection{Emu CMB accuracy}
\begin{figure}
  \centerline{
    \plotone{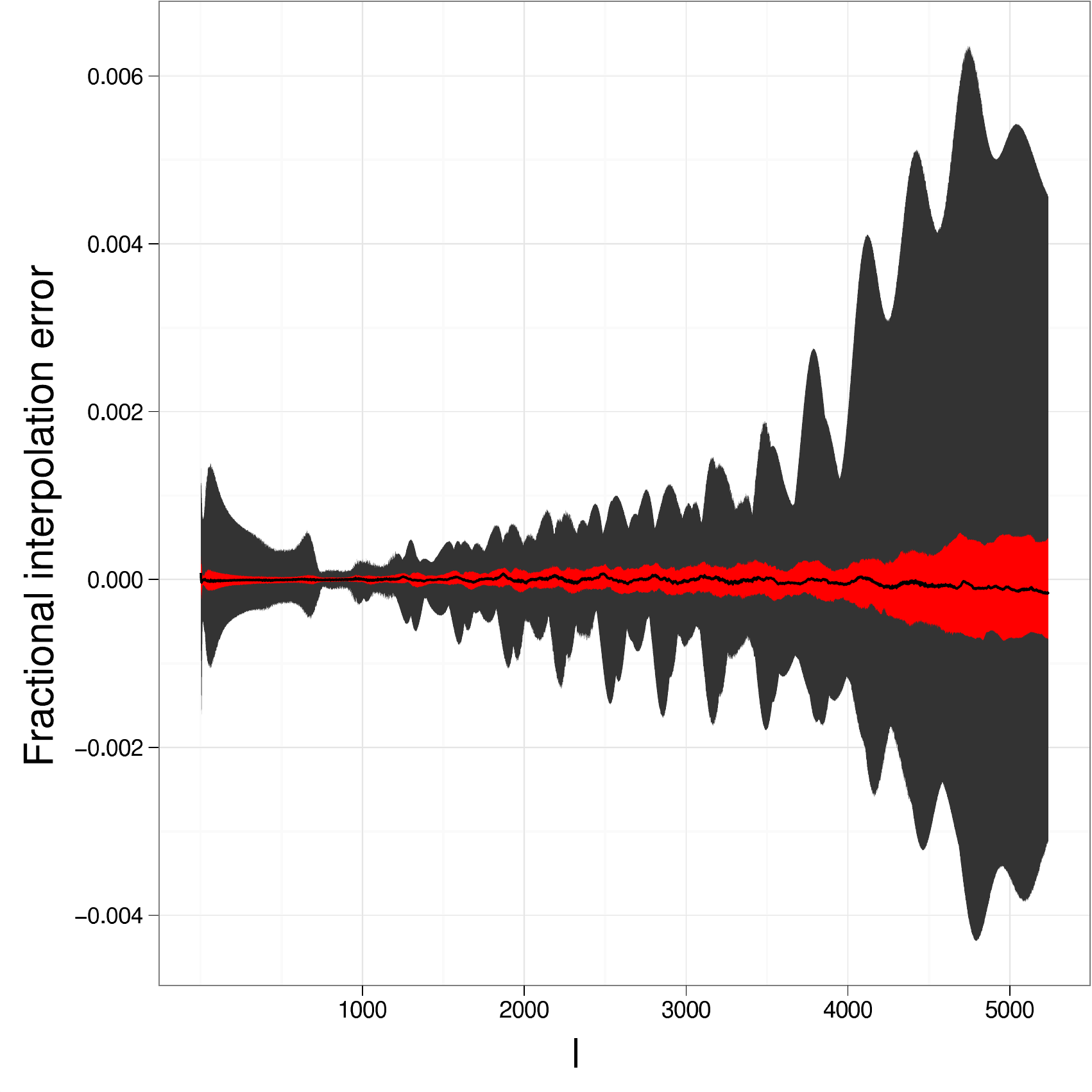}
  }
  \caption{\label{fg:emuInterpErrors} Fractional errors in the
    interpolation of the power spectra for the \textsc{Emu CMB} design.
    See Figure~\ref{fg:deltaCl} caption for the meaning of the colors.}
\end{figure}

We validated the emulator by running \textsc{CAMB} at an additional 100 points
in a new realization of an OALHSFS design.  (Note the validation
  plots in the main body of the paper use points from a PS design.)
The interpolation errors for the spectra at these 100 test points are
shown in Figure~\ref{fg:emuInterpErrors}.  For all test points, the emulator
is accurate to less than 0.6\%  out to $\ell=5238$.  The errors are
less than 0.2\% for $\ell < 3000$, which is consistent with Figure~\ref{fg:deltaCl}.

Note that while \textsc{Emu CMB} uses significantly fewer training runs of \textsc{CAMB}
than \textsc{PICO}, unlike \textsc{PICO} this \textsc{Emu CMB} example does \emph{not} predict
polarization power spectra.  We have verified with simple tests that
interpolation of the low-$\ell$ parts of the EE and TE CMB power
spectra is more challenging than the TT spectra demonstrated
here.  In addition, many of the training runs for
\textsc{PICO} were required to include cosmological models with nonzero spatial
curvature, which we have not considered\footnote{C. Fendt private communication.}.

\section{Matter power spectrum}\label{sec:coyoteUniverse}
In this section we compare the polynomial interpolation errors for OALHS designs for
the dark matter power spectrum to the GP interpolations demonstrated in \citet{heitmann09}.  Following
\citet{heitmann09} we use the \textsc{HALOFIT}~\citep{smith03} algorithm as
provided in \textsc{CAMB}~\citep{camb} to generate nonlinear dark matter power
spectra as functions of the five cosmological parameters: $\Omega_c
h^2$, $\Omega_b h^2$, $n_s$, $w$, $\sigma_8$.  The reduced Hubble
parameter $h$ is determined from $\omega_c$ and $\omega_b$ assuming
spatial flatness and a fixed angular size of the sound horizon at last
scattering \citep[$\theta_{*} = \pi/302.4$ steradians as defined
in][]{heitmann09}.  The prior parameter ranges defining the boundaries
of the OALHS design are the same as \citet{heitmann09}.

\begin{figure}[ht]
  \centerline{
    \plotone{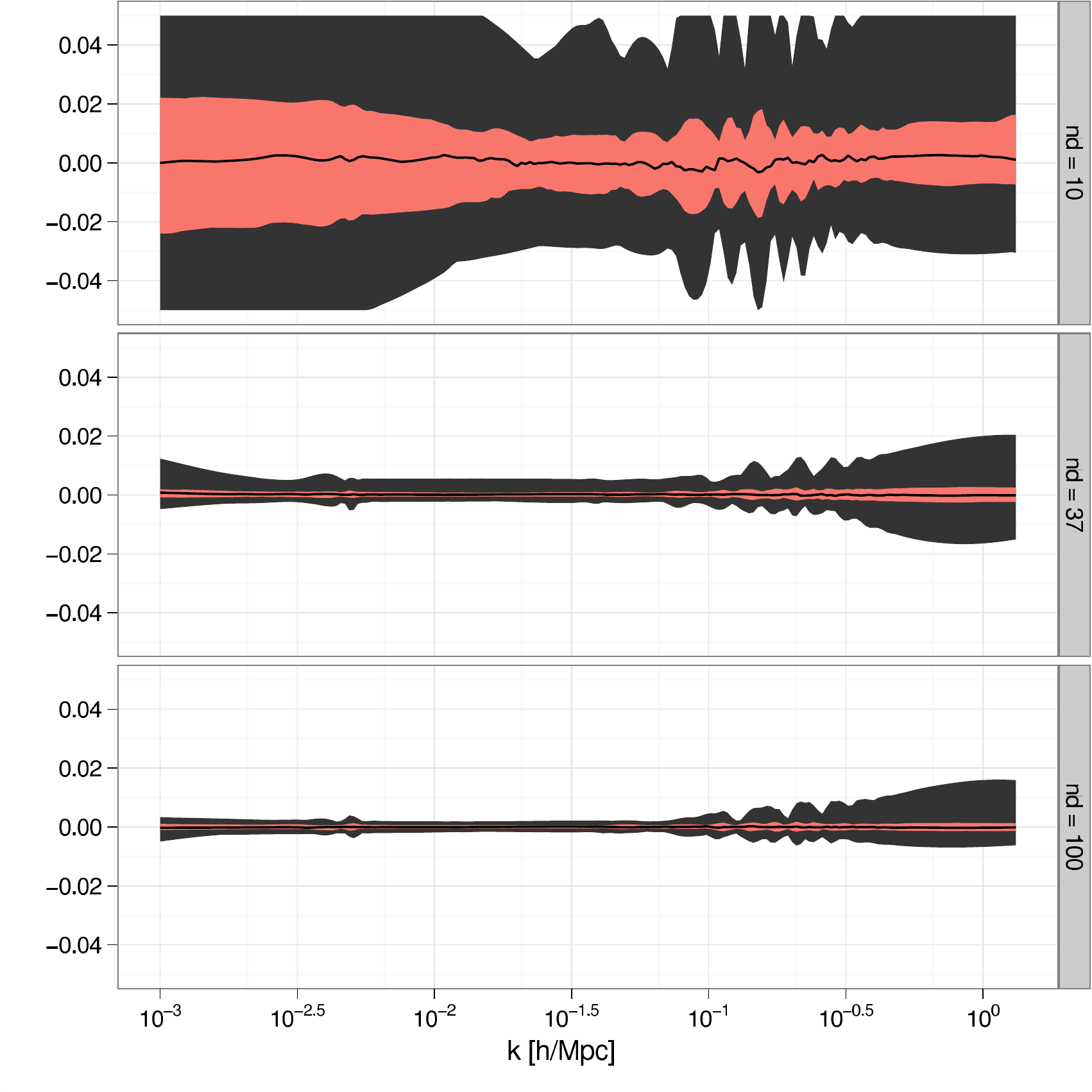}
 }
  \caption{\label{fg:coyoteInterpErrors}Fractional error in the
    interpolated matter power spectra using polynomial interpolation
    for 3 OALHS designs with
    different $n_d$.  The parameters and widths of the OALHS designs 
    match those in~\citet{heitmann09}.  For $n_d=10$ the mode
    amplitudes in the decomposition of the power spectra are fit with
    linear polynomials.  For $n_d=37,100$ the mode amplitudes are fit
    with quadratic polynomials.}
\end{figure}

The fractional interpolation errors for three designs with
$n_d=10,37,100$ are shown in Figure~\ref{fg:coyoteInterpErrors}.
\citet{heitmann09} built an OALHS design with $n_d=37$ and
demonstrated sub-percent interpolation errors using GP interpolation
over  the same range in
wavenumber as shown in Figure~\ref{fg:coyoteInterpErrors}.  Using
quadratic polynomial interpolation, the errors are worse, but are
still under 2\% for all wavenumbers.  Increasing $n_d$ to 100 points
reduces the interpolation errors to sub-percent levels for all but a
few test points in the design.  
For the matter power spectrum over this design space, the GP
interpolation developed by~\citet{heitmann09} therefore requires three
times fewer design points to achieve the same accuracy as a naive
quadratic polynomial fit.  However, we note that the OALHSFS design
demonstrated in Section~\ref{sub:designs_compared} 
can achieve the same reduction in the required number of design points
to reach a fixed accuracy while using the much simpler polynomial
interpolation.  

We take these results as confirmation that the good performance of
polynomial interpolation in the CMB designs in the main body of the
paper is not limited to a single physical example.  More generally, the success
of a simulation emulator depends strongly on the simulation design
(including the form of the mode decomposition of the power spectrum)
while optimized GP interpolation can also significantly improve the
performance of a given Latin-hypercube-based design (in contrast to a
PS-based design).
\end{document}